\documentclass[preprint,5p,review,times,twocolumn]{elsarticle}

\usepackage{amssymb}
\usepackage{amsmath}

\usepackage{booktabs}
\usepackage{caption}
\usepackage{subcaption}
\usepackage{subfloat}
\usepackage{lineno}

\begin{document}

\begin{frontmatter}



\title{Extreme Value Estimates using Vibration Energy Harvesting}


\author[add1,add2]{George Vathakkattil Joseph \corref{cor1}}
\ead{george.vathakkattiljoseph@ucdconnect.ie}
\author[add3,add4]{Guangbo Hao}
\author[add1,add2]{Vikram Pakrashi }
\address[add1]{Dynamical Systems and Risk Laboratory, University College Dublin}
\address[add2]{Marine and Renewable Energy Ireland (MaREI) Centre, University College Dublin}
\address[add3]{School of Engineering, University College Cork}
\address[add4]{Marine and Renewable Energy Ireland (MaREI) Centre, Environmental Research Institute, University College Cork}
\cortext[cor1]{Corresponding author}

\begin{abstract}
This paper establishes the possibility of utilising energy harvesting from mechanical vibrations to estimate extreme value responses of the host structure and demonstrates the calibration of these estimates for excitation spectra typical to natural systems. For illustrative purposes, a cantilever type energy harvester is considered for wind excitation. The extreme value estimates are established through a Generalised Pareto Distribution (GPD). Classically well-known Kaimal and Davenport spectra for wind have been considered in this paper for comparison purposes. The work also demonstrates how return levels can be mapped using energy harvesting levels and indicates that vibration energy harvesting, in its  own right has the potential to be used for extreme value analysis and estimates. The work has impact on health monitoring and assessment of built infrastructure in various stages of repair or disrepair and exposed to nature throughout their lifetime.
\end{abstract}


\end{frontmatter}


\section{Introduction}
\label{}
Estimation of extreme events and their responses is crucial in designing and engineering structures that can withstand the forces of the physical environment they are subjected to. This is particularly relevant both for ageing built infrastructure\cite{znidaric2011review} burgeoning sectors\cite{quilligan2012fragility}. The relationship between climate variability or change estimates with those of extreme values\cite{cooley2009extreme} makes such estimates more relevant. Estimating the probability of occurrence of severe natural phenomena is of importance to any form of long-term planning. Engineering bridges, dams, seawalls, off-shore structures, windmills, skyscrapers, etc. are dependent on such estimation and therefore assessment of extreme dynamic responses has been a popular subject of study. The statistical methods of extreme value theory have been developed to facilitate this essential requirement of estimating the probability of extreme levels of a process based on previously observed data. Extreme values are usually characterised as return periods, which is a derived representation from the statistical properties of the processes. There have been numerous studies that condensed empirical data pertaining to several natural phenomena such as wind, ocean waves, and earthquakes to a set of spectra\cite{pierson1964proposed,elfouhaily1997unified,brune1970tectonic,kaimal1972spectral,davenport1961spectrum}, ie. the power spectral densities (PSD), and the probability distribution functions (PDF), \cite{morgan2011probability,carta2009review,huang1980experimental,thornton1983transformation,forristall1978statistical}, using which the processes can be approximated.  

The dynamics of a wide range of structures and their operational conditions can be approximated by using an assumption of linearity and with a single degree of freedom (SDOF) dominating. Recently, with the rise of renewable energy devices (onshore and offshore wind turbines, wave devices, combined wind and wave devices etc.), the structures have become light and more flexible\cite{arrigan2011control}. While this leads to the complexities of nonlinearity or the need to understand more than one degree of freedom, the fundamental basis behind several such existing or conceptual designs at various technological readiness levels are often adequately represented by SDOF systems. This consideration highlights the importance of accurate estimation of return periods. New designs can be guided by the estimates of extreme values and this is reflected in their fragility curves. Studying the dynamic responses of the structures to extreme levels can be used to assess structural non-linearities as well\cite{pakrashi2016assessment}. Accurate estimation of the change in extreme value distributions would also contribute to the quantification of change in operational environment or the system itself.

The accuracy of the estimation relies on having large data-sets, typically over many years, with good resolution. Also, the non-stationary nature of physical phenomena warrant perpetual monitoring of the events, to update the estimates periodically. Conventional methods of measurement like strain gauges, anemometers, inertial measurement units placed in buoys deployed in the ocean, etc. require power sources to function. This may not be feasible in many cases and a battery may typically requires replacement or a solar/wind powered recharging setup for extending it's lifetime for long-term monitoring. Under such circumstances, technologies and methods for extreme value estimates of the responses of the built infrastructure requiring low to no power can be very attractive. 

Vibration energy harvesters present themselves as a promising solution for powering such sensors and this has been studied extensively by several groups\cite{li2014energy, siddique2015comprehensive}. Vibration energy harvesters operate by responding to the dynamic response signatures of the host device. For example, dynamic responses of structures can be the base excitation of an energy harvester, which then is converted to voltage through electromechanical coupling and a circuit, which may be as simple as a resistance or more complex, based on the requirements of the final application. Under such circumstances, there is an opportunity to use energy harvesters in their own right as monitors. Energy harvesting based structural health monitoring (SHM) has been recently suggested\cite{cahill2016effect,cahill2014energy}. Such use also reduces instrumentation and helps site-safety in locations where access or risk can be significant\cite{pakrashi2012monitoring}. Another advantage lies in the fact that the magnitude of harvested energy from built infrastructure tends to be small and thus a physical and useful interpretation of its variation, rather than its absolute magnitude, can be more rewarding. In this paper, we demonstrate that the voltage signals generated from a vibration energy harvester subjected to typical natural systems, can be used to estimate extreme value responses of such systems. We show that extreme value responses estimated using inertial data directly correspond to the estimation using voltage signals from the harvester. 

\section{Conceptual background}
The most common approaches to extreme value estimation are based on asymptotic distributions\cite{coles2001introduction}. One approach is to assume that the epochal extremes, such as annual or monthly maxima, are distributed according to the Generalised Extreme Value (GEV) distribution, and estimating the parameters of the distribution based on the observed data. This is referred to as the block-maxima approach\cite{coles2001introduction}. A second approach utilising all available data instead of just sparse maxima, is to assume that the exceedances above a threshold follow a a Generalised Pareto distribution (GPD) and estimate the model parameters fitting the data\cite{coles2001introduction}. Since we are not theoretically limited in resolution while performing a numerical analysis, and since a better utilisation of the data can be made by using the entire time-series\cite{coles2001introduction}, we employ the second approach, called the peaks-over-threshold method, for our estimates. For a large enough threshold $u$, the distribution function of $(X-u)$, conditional on $X > u$, referred to as GPD is:

\begin{linenomath*}
\begin{equation}
\label{GPD}
H(X-u) = 1 - \left(1+\frac{\xi (X-u)}{\sigma}\right)^{-1/\xi}
\end{equation}
\end{linenomath*}

where $\xi$ called the shape parameter and $\sigma$ as the scale parameter. The parameters can be estimated from the data using the maximum likelihood estimation method\cite{coles2001introduction}.

Piezoelectric vibrational energy harvesting has been modelled previously with varying levels of complexity and accuracy\cite{williams1996analysis,dutoit2005design,Erturk2008, erturk2008distributed}. As the intention of this paper is to provide a proof of concept, we settle for a reasonable and established model using the corrected lumped-parameter approach from Erturk and Inman \cite{Erturk2008}. A cantilever in bi-morph configuration is modelled. The electromechanically coupled equations of a piezoelectric vibration energy harvester are:

\begin{linenomath*}
\begin{equation}
\label{coupled1}
     m_h\ddot{z} +  c_h\dot{z} + k_hz - \theta V = -\mu m_h\ddot{y}
\end{equation}
\end{linenomath*}

\begin{linenomath*}
\begin{equation}
\label{coupled2}
     \theta \dot{z} + C_p\dot{V} + \frac{1}{R_l}V = 0
\end{equation}
\end{linenomath*}

where $m_h$ is the (Rayleigh) equivalent mass of the harvester, $c_h$ is the damping, $k_h$ is the stiffness, $y$ is the base excitation of the harvester, $\theta$ is the electromechanical coupling coefficient, $V$ is the voltage across the piezoceramic, $C_p$ is the capacitance of the piezoceramic and $R_l$ is the load resistance. $\mu$ is a correction factor approximately given by 

\begin{linenomath*}
\begin{equation}
\mu = \frac{(M_t/m)^2 + 0.603(M_t/m) + 0.08955}{(M_t/m)^2 + 0.4637(M_t/m) + 0.05718}
\end{equation}
\end{linenomath*} 

where $M_t$ is the tip mass and $m$ is the beam mass of the harvester.
The parameters used are given in Table 1. 

Natural systems such as ocean waves, wind, etc. can be approximately represented by specific empirically derived spectra. These spectra can be used to generate time-series ensembles representing the system. A common approach to generate time-series would be to apply an inverse Fourier transform, but the Fourier coefficients of the signals are not directly related to the spectrum and thus have to be generated randomly. The standard deviation of each coefficient is determined by the spectrum, given by: 

\begin{linenomath*}
\begin{equation}
\label{std_dev}
\sigma^2_X = \frac{T}{2\pi}S(\omega)
\end{equation}
\end{linenomath*}

where $S(\omega)$ is the spectrum and $T$ is the total sampling duration.
A Monte-Carlo simulation employing randomly generated coefficients yields time-series ensembles representing the phenomena. The probability density function (PDF) of the ensemble can then be used to estimate extreme values by fitting the tail of the distribution to a standard extreme value distribution using the approaches described above.

\begin{figure}[t]
\centering
  \includegraphics[width=1.00\linewidth]{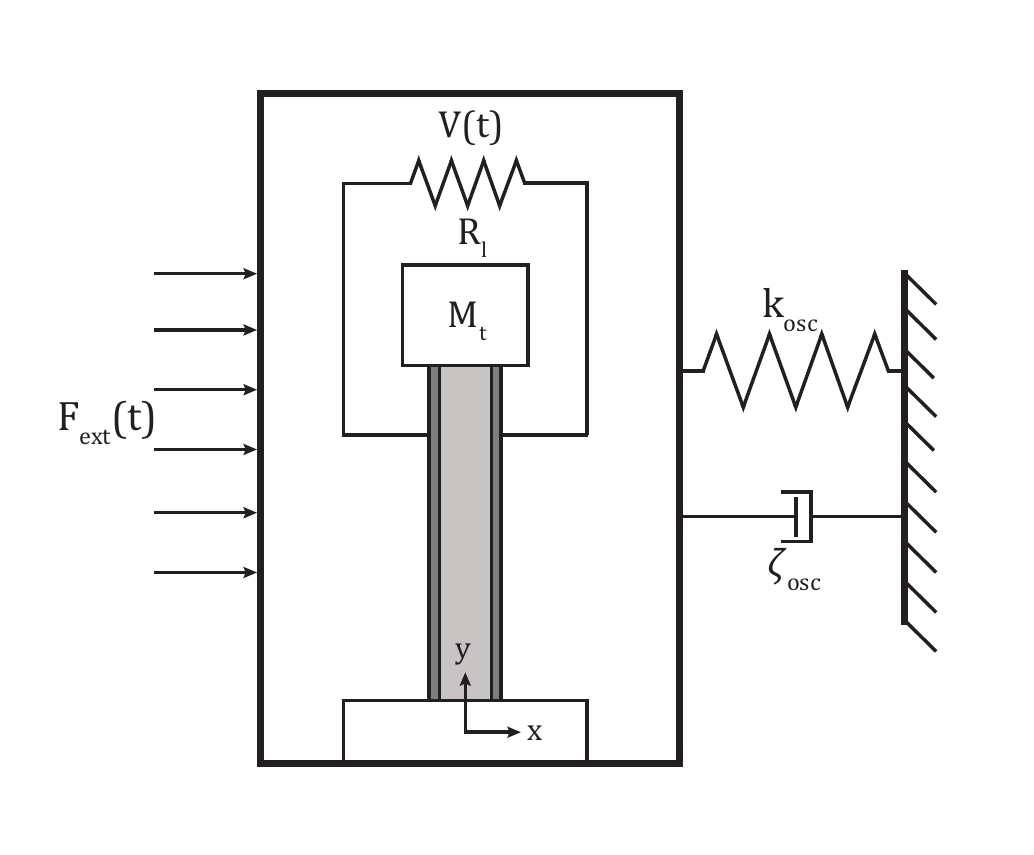}
  \caption{Schematic of the harvester system}
  \label{schematic}
\end{figure}

The dynamic response of the structure under analysis, excited by the process represented by the time-series, can be expressed as the output obtained by passing an input signal through a system. Considering the structure to be a harmonic oscillator,

\begin{linenomath*}
\begin{equation}
\ddot{x}+2\zeta_{osc} \omega_{osc}\dot{x}+\omega_{osc}^2x=F_{ext}/M_{osc}
\end{equation} 
\end{linenomath*}

where $\omega_{osc}$ is the natural frequency, $\zeta_{osc}$ is the damping ratio, and $M_{osc}$ is the mass of the SDOF system, we have a second order linear system through which the excitation time-series, $F_{ext}$, representing a realisation of the process is passed through. Since the structure hosts the energy harvester, the acceleration of the structure acts as the base acceleration of the harvester (ref Fig. \ref{schematic}). In the case of a harvester, for a linear stochastic process the power spectral density(PSD) of the base acceleration is related to the PSD of the harvester voltage through the voltage frequency response function (FRF) $\alpha(\omega)$ obtained from harmonic excitation\cite{erturk2011piezoelectric}

\begin{linenomath*}
\begin{equation}
\label{PSDrelation}
    S_v(\omega) = |\alpha(\omega)|^2S_a(\omega)
\end{equation} 
\end{linenomath*}


Thus we have a linear system again between the acceleration from the host structure and the harvester output. Since the cascaded transformations are linear, the cumulative distribution function (CDF) of the voltage generated can be mapped to the CDF of the base-acceleration of the harvester and also to the CDF of the excitation process. This establishes that any extreme value estimation done purely based on the spectra, or through the data obtained from a traditional inertial sensor can be done with the same approach and confidence level using an energy harvester, once calibrated. 


\section{Numerical Analysis}

A numerical experiment was carried out to estimate extreme values based on a) the excitation time-series, b) acceleration of the structure, and c) voltage generated by the harvester, which were then compared. As a benchmark analysis, the host structure was excited using Gaussian white noise as the forcing function. To demonstrate the applicability in a practical scenario, two classical spectra for wind speed were chosen next, Kaimal spectrum and Davenport spectrum. Assuming stationarity and ergodicity of the process, an ensemble of time-series generated randomly based on the theory mentioned above, captures the statistics of the process and extreme value theory can be applied to the combined data set. Stationarity of the process is assumed for the sake of simplicity and is not pre-requisite as the mapping is between return levels (shown in section 4). Extreme value methods applied to non-stationary processes can be applied in the same sense. 

For Gaussian white noise, a time-series of power 30dBW was generated. For the wind-speed spectra,  Fourier coefficients were generated, where each coefficient is generated using a normal distribution of mean 0 and standard deviation $\sigma_X$ (Eq.\ref{std_dev}). The first Fourier coefficient corresponds to 0 frequency and determines the mean of the time series, and is given by $\frac{TU}{2\pi}$. For an accurate inverse transform, complex numbers are used for the Fourier coefficients with the standard deviation distributed equally among the real and imaginary parts

\begin{linenomath*}
\begin{equation}
\sigma_{a_n}^2 = \sigma_{b_n}^2 = \frac{1}{2}\sigma_{X_n}^2
\end{equation} 
\end{linenomath*}



For wind speed, the spectra are defined for a given mean wind speed $U(z)$ at a given height $z$. The Kaimal spectrum\cite{kaimal1972spectral} is expressed as:

\begin{linenomath*}
\begin{equation}
\frac{nS(n)}{u_f^2} = \frac{105f}{\left(1+33f\right)^{5/3}} 
\end{equation}
\end{linenomath*}

where $n$ is the frequency in Hertz, $S(n)$ is the power spectral density of wind speed fluctuations, $f$ is the normalised frequency $nz/U(z)$, and $u_f$ is the friction velocity given by 

\begin{linenomath*}
\begin{equation}
u_f = \frac{kU(z)}{ln(z/z_0)}
\end{equation}
\end{linenomath*}

where $k$ is the von Karman's constant ($k$ = 0.4) and $z_0$ is the reference height ($z_0$ = 0.025).

The Davenport spectrum\cite{davenport1961spectrum} is expressed as:

\begin{linenomath*}
\begin{equation}
\frac{nS(n)}{u_f^2} = \frac{4x^2}{\left(1+x^2\right)^{4/3}} 
\end{equation}
\end{linenomath*}

where $x = 1200f/z$.
The wind velocity time series is used to calculate pressure by utilising the simple relation 

\begin{linenomath*}
\begin{equation}
p = \frac{1}{2}\times \text{density of air} \times \text{wind velocity}^2\times \text{shape factor}
\end{equation}
\end{linenomath*}

where density of air is taken as 1.25 and shape factor as unity. Force is calculated by multiplying with unit area. The force acts upon the host structure and thus the force time series is applied to a harmonic SDOF oscillator whose parameters are listed in Table \ref{params}. The energy harvester is attached to the host structure at its base, thereby making the base acceleration of the harvester the same as the acceleration of the SDOF oscillator. 
The parameters of the harvester(from \cite{cahill2016effect}) are listed in Table \ref{params}. The second order equation for the SDOF oscillator and the coupled equations for the harvester are solved at the two stages based on an explicit Runge-Kutta (4,5) scheme, the Dormand-Prince pair\cite{dormand1980family}. The time-series is generated at a sampling frequency of 25Hz for a duration of 3600s leading to 90,000 points per series. An ensemble of 50 series is generated for each spectrum. After obtaining the time-series of the wind-speed, the acceleration, and the voltage, we proceed to estimate extreme values using each data-set.

\begin{table}[]
\centering
\begin{tabular}{@{}lll@{}}
\toprule
Parameter                & Value    & Unit \\ \midrule
\textbf{SDOF oscillator} &         &      \\
Mass                     & 1       & kg   \\
Damping ratio            & 0.02148 &      \\
Natural frequency (tuned)        & 12.79   & Hz   \\
\textbf{Harvester}       &         &      \\
Tip mass                 & 0.03    & kg   \\
Beam mass                & 0.01365 & kg   \\ 
Damping ratio            & 0.04     &       \\
Load resistance          & 1000000    & $\Omega$   \\
Capacitance of piezoceramic     & 1.966  & nF   \\
Natural frequency        & 12.79    & Hz   \\
Electromechanical coupling & 1.289    & $\mu$C/m  \\\bottomrule
\end{tabular}
\caption{Harvester and oscillator parameters}
\label{params}
\end{table}

In order to fit a GPD model, an appropriate threshold choice should be made. A reasonably high threshold can be chosen in many cases from physical interpretation of the data. A practical engineering approach to choose a threshold would be to look at the empirical CDF of the data and choose a high enough threshold below the required percentile. The percentiles above which extremes are to be estimated are generally above the 95th percentile, which correspond to a CDF value of 0.95. The empirical cumulative distribution function of the absolute values of each data-set is shown in Fig. \ref{}. We have considered the magnitude of acceleration and voltage as the it is often more important for dynamic responses. Using the value at CDF = 0.95 as the threshold, we proceed to fit the distribution using the maximum likelihood estimation method and find the model parameters. 

\section{Results}

The thresholds used and the model parameters obtained for each time-series are listed in Table \ref{fittable}. The data is trimmed by 10 seconds at the beginning to ensure stationarity. Plots showing goodness of fit and the CDF of the Gaussian white noise process is given in Figures \ref{CDF_combined_white}-\ref{white_accel_probplot}. The voltage levels in the CDF show good agreement with the results derived in \cite{erturk2011piezoelectric}. The fitted distribution probability plots are coincident across most of the plot showing the validity of using the GP distribution and the estimation method we adopted. 

\begin{table}[]
\centering
\begin{tabular}{@{}lll@{}}
\toprule
Parameters                        & Kaimal    & Davenport \\ \midrule
\textbf{Threshold}                &         &      \\
Wind-speed (\textit{ms}\textsuperscript{-1})	  & 24.75   &  25.36  \\
Acceleration (\textit{ms}\textsuperscript{-2})             & 19.01   & 13.97   \\
Voltage (\textit{V})              & 2.474    &   1.819     \\
\textbf{GPD fit}                  &          &          \\
Shape parameter - Wind-speed      & -0.137    &   -0.177   \\ 
Shape parameter - Acceleration    & -0.073    &   -0.073   \\ 
Shape parameter - Voltage         & -0.074    &   -0.075   \\
Scale parameter - Wind-speed      & 1.356    &   1.522   \\
Scale parameter - Acceleration    & 4.320    &   3.271   \\
Scale parameter - Voltage         & 0.562    &   0.427    \\ \bottomrule
\end{tabular}
\caption{Thresholds and fitted model parameters}
\label{fittable}
\end{table}

\begin{subfigures}
\begin{figure}[]
	\centering
	\includegraphics[width=1.00\linewidth]{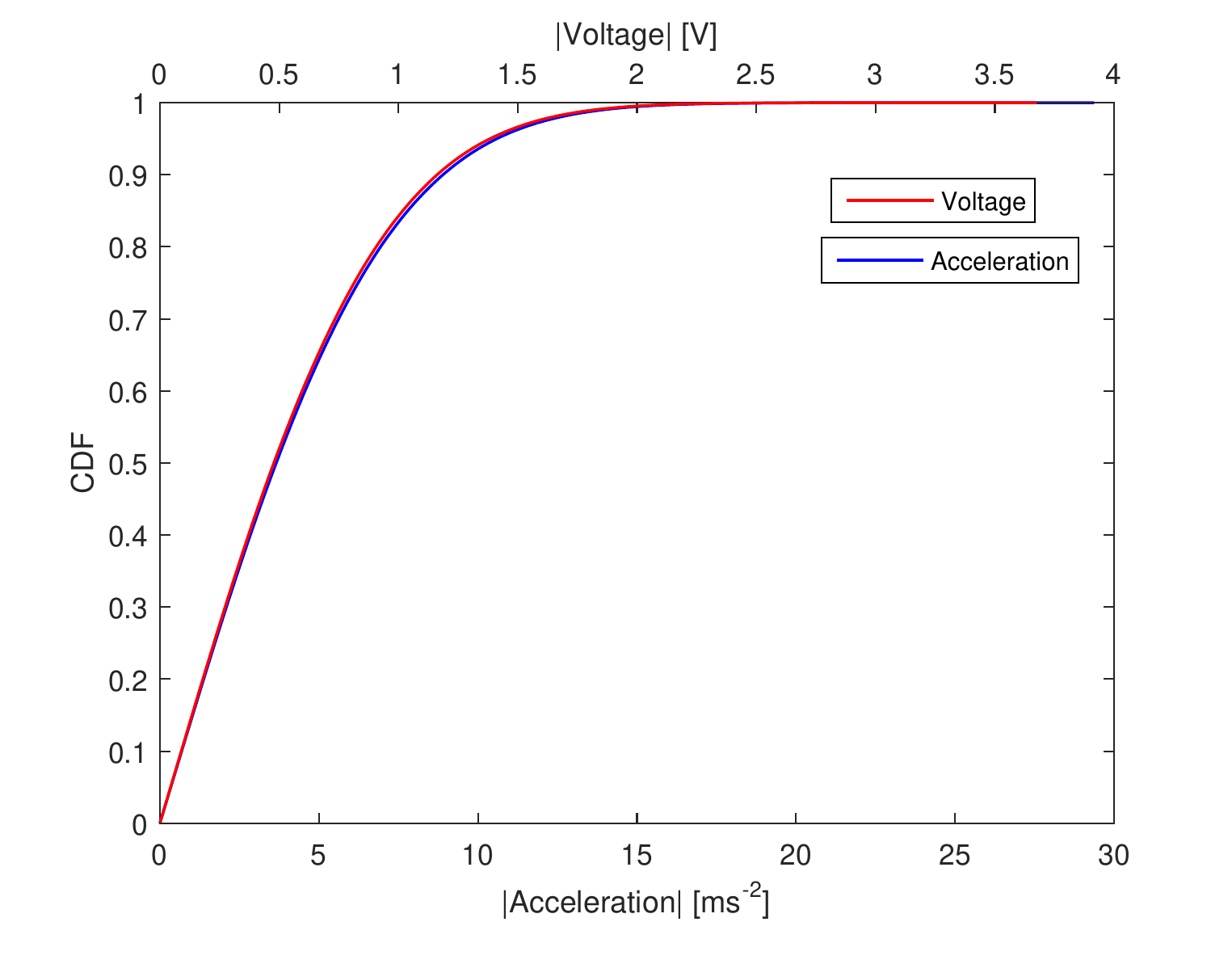}
	\caption{Gaussian white noise empirical CDF}
	\label{CDF_combined_white}
\end{figure}

\begin{figure}[]
	\centering
	\includegraphics[width=1.00\linewidth]{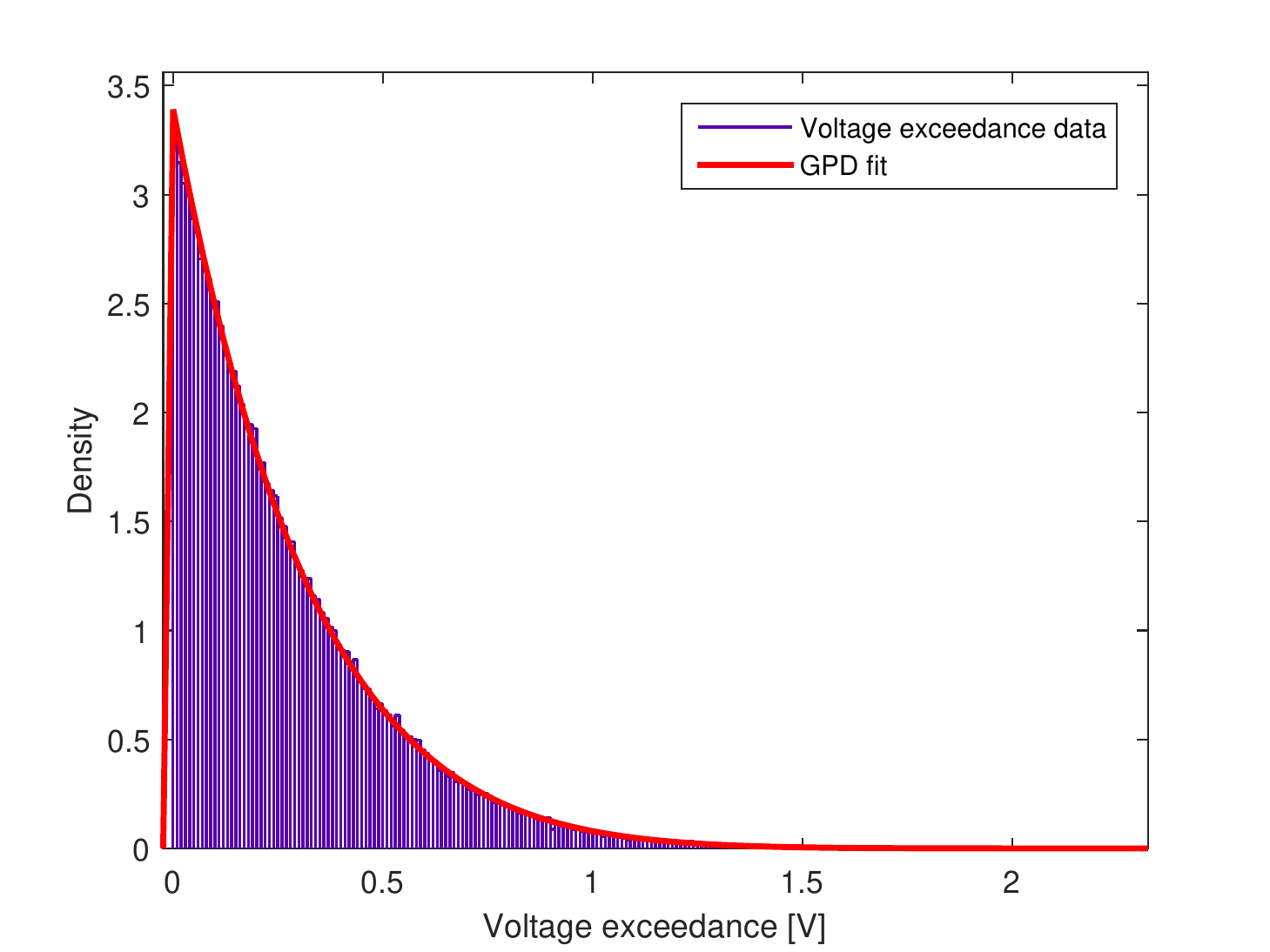}
	\caption{GWN voltage threshold exceedance density plot}
	\label{white_volt_PDF}
\end{figure}

\begin{figure}[]
	\centering
	\includegraphics[width=1.00\linewidth]{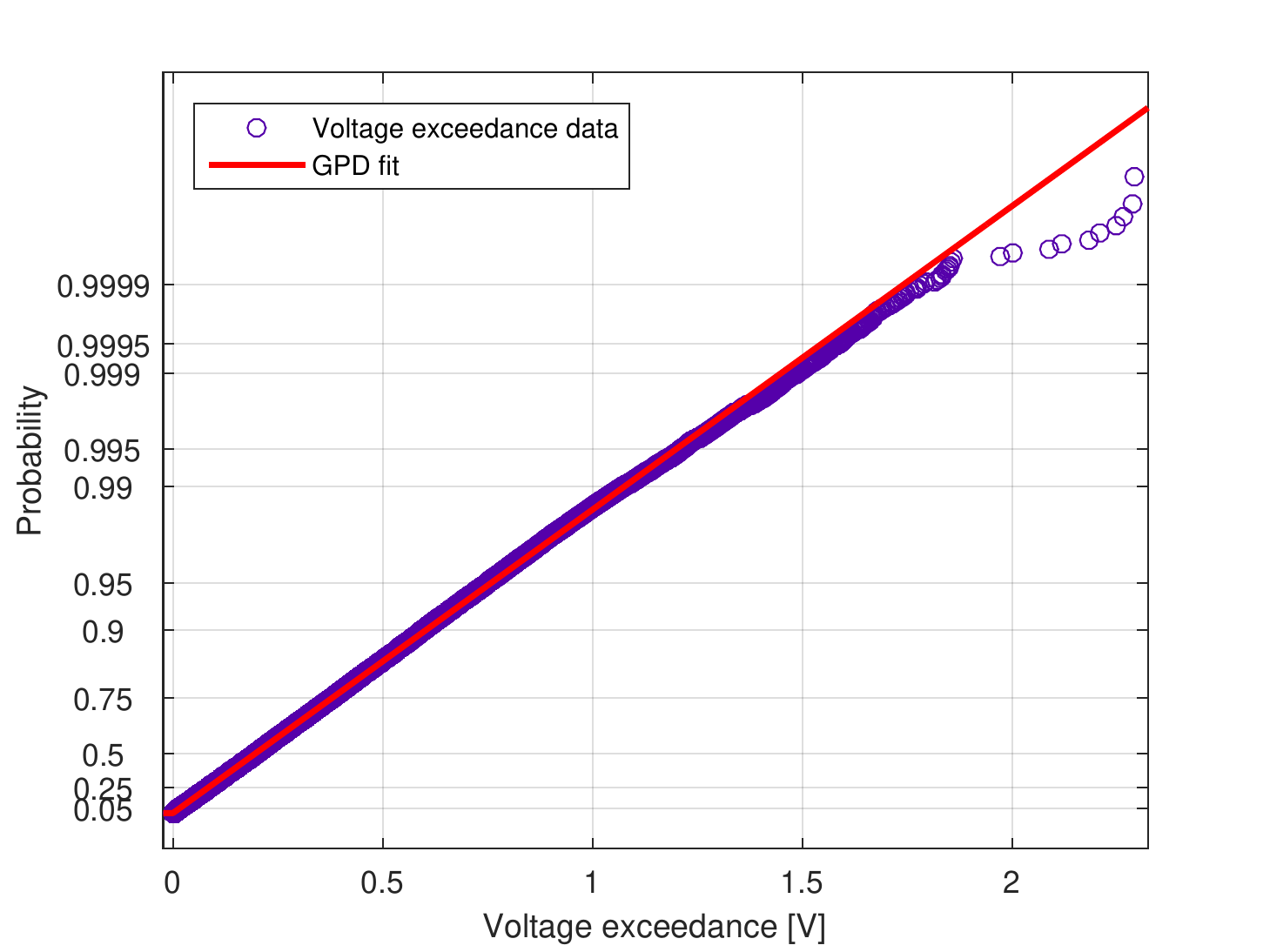}
	\caption{GWN voltage threshold exceedance probability plot}
	\label{white_volt_probplot}
\end{figure}

\begin{figure}[]
	\centering
	\includegraphics[width=1.00\linewidth]{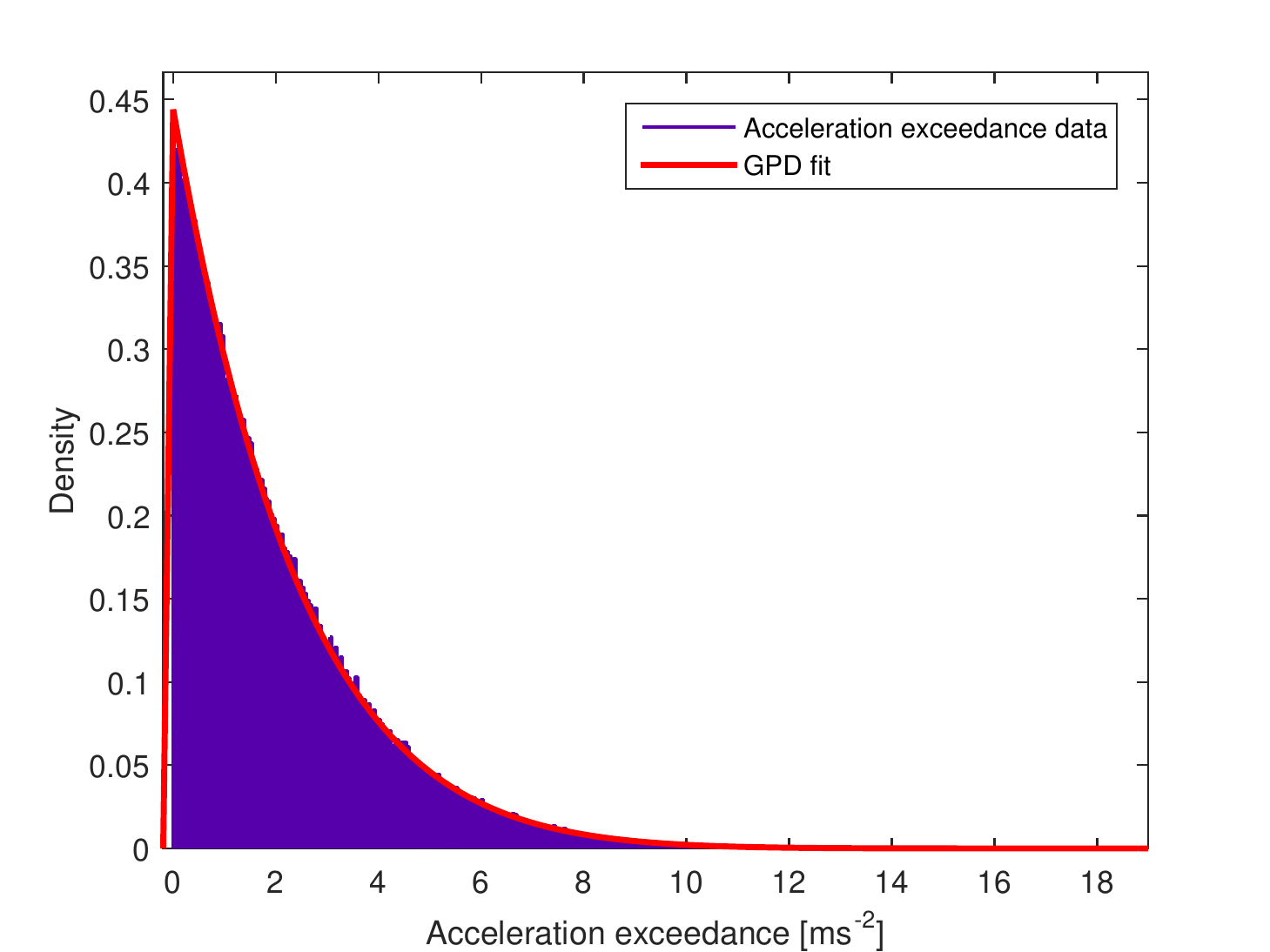}
	\caption{GWN acceleration threshold exceedance density plot}
	\label{white_accel_PDF}
\end{figure}

\begin{figure}[]
	\centering
	\includegraphics[width=1.00\linewidth]{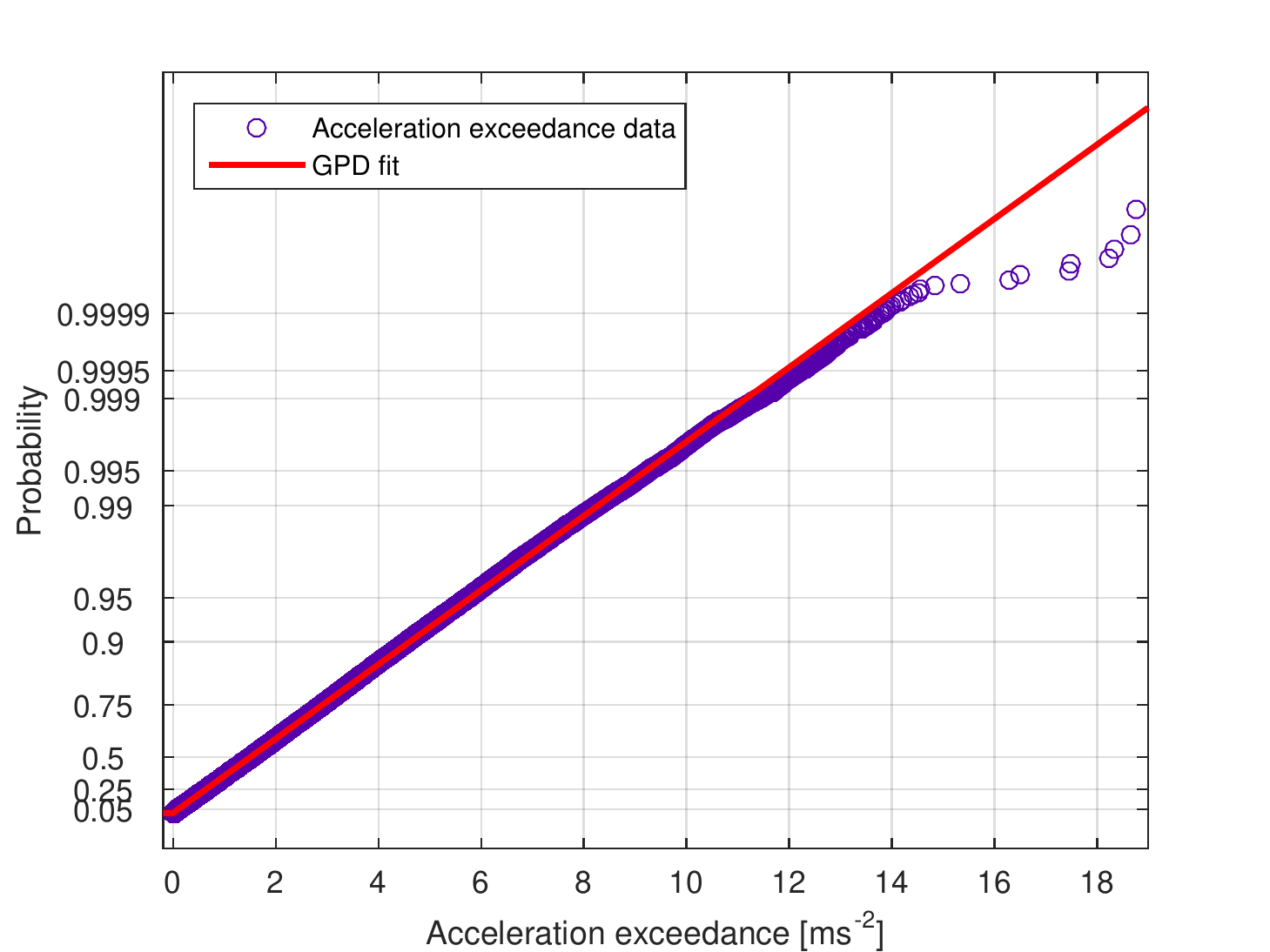}
	\caption{GWN acceleration threshold exceedance probability plot}
	\label{white_accel_probplot}
\end{figure}
\end{subfigures}
Figures \ref{CDF_combined_kaimal}-\ref{davenport_accel_probplot} correspond to the wind-spectra. The probability plots show that the fits are reasonably good in these cases as well. It may be noted here that the token threshold choice of 0.95 can be altered for a better fit as required by the estimation case and guided by the threshold selection constraints imposed by the data \cite{coles2001introduction}. 

\begin{subfigures}
\begin{figure}[]
	\centering
	\includegraphics[width=1.00\linewidth]{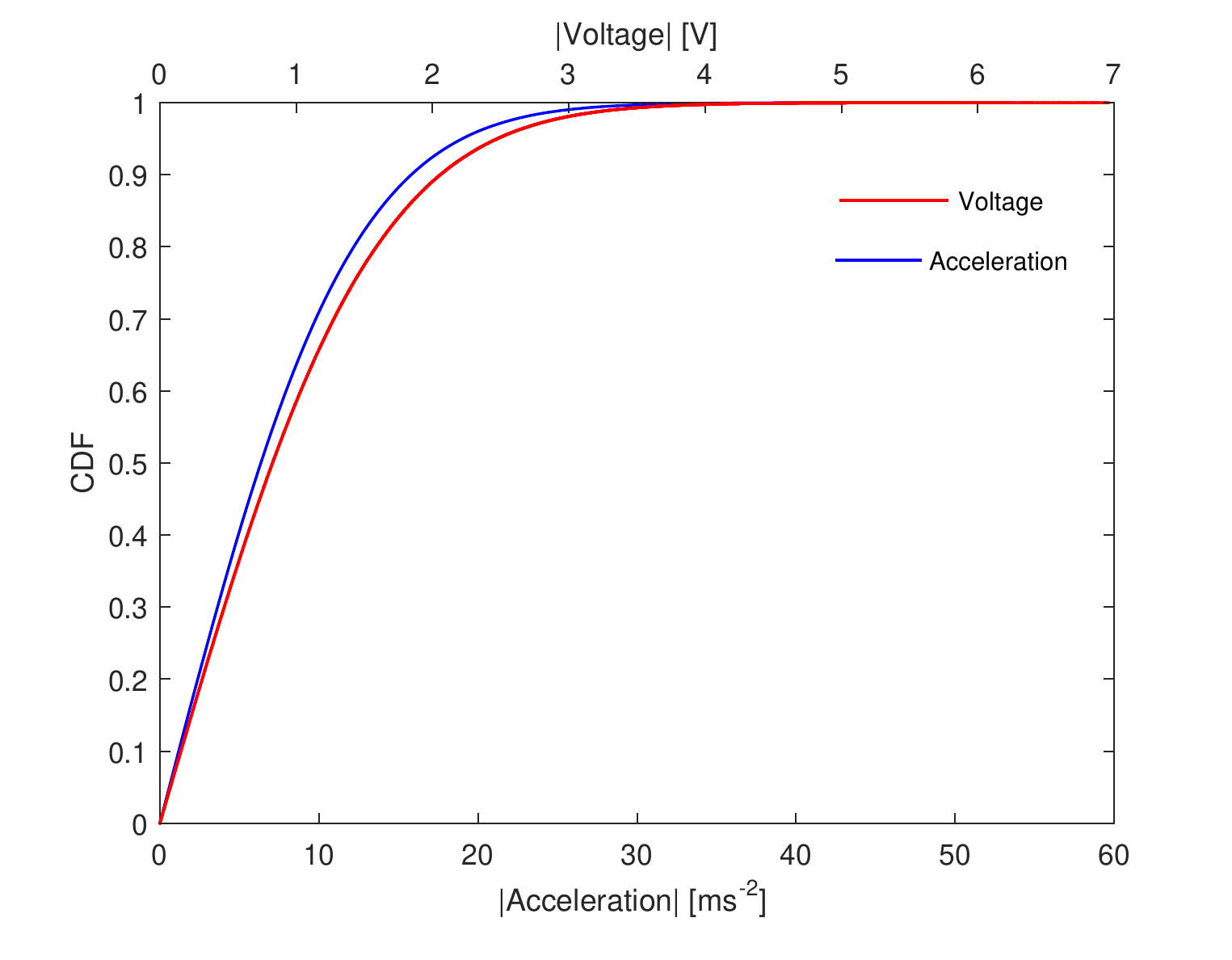}
	\caption{Kaimal empirical CDF}
	\label{CDF_combined_kaimal}
\end{figure}

\begin{figure}[]
	\centering
	\includegraphics[width=1.00\linewidth]{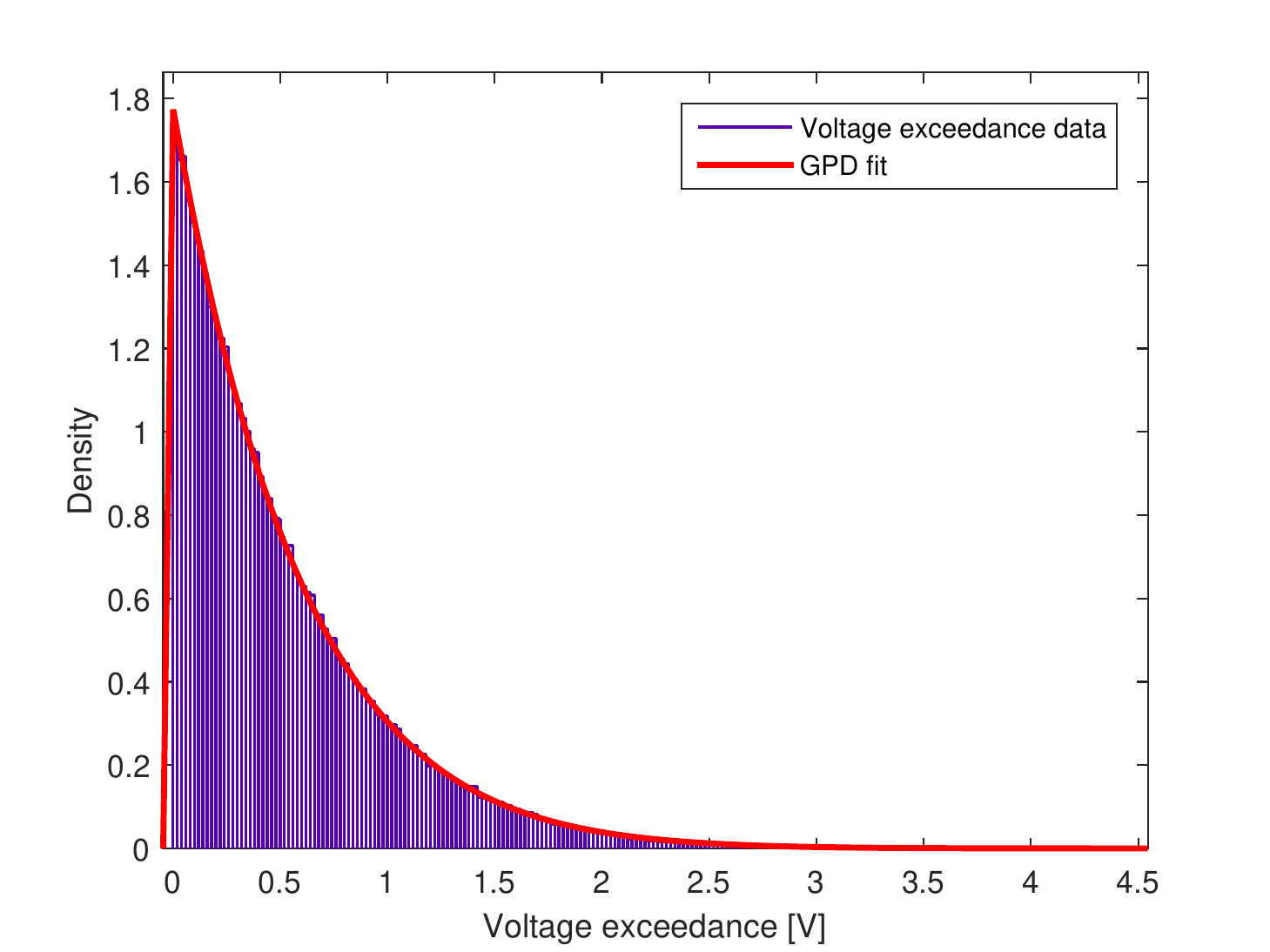}
	\caption{Kaimal voltage threshold exceedance density plot}
	\label{kaimal_volt_PDF}
\end{figure}

\begin{figure}[]
	\centering
	\includegraphics[width=1.00\linewidth]{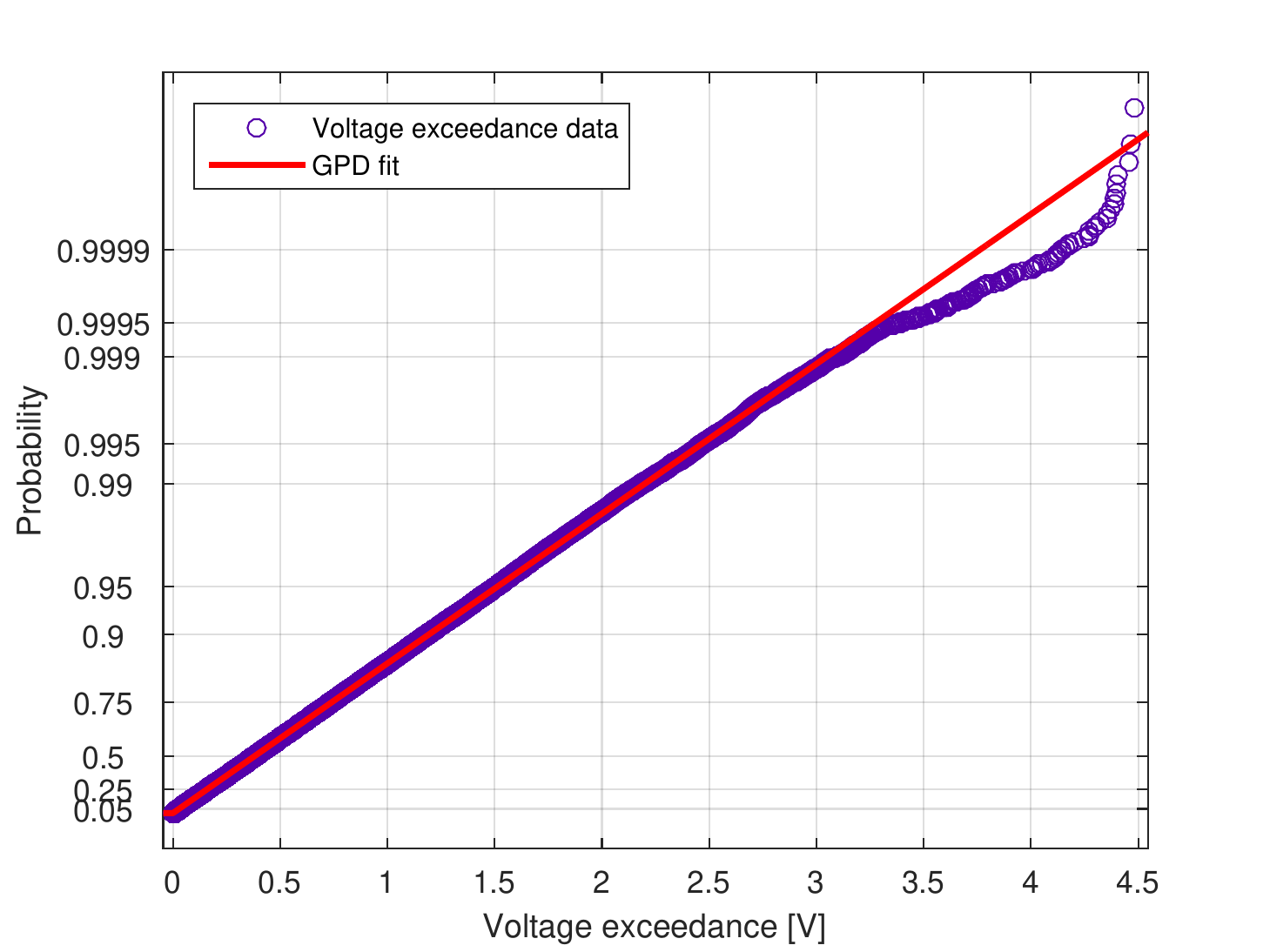}
	\caption{Kaimal voltage threshold exceedance probability plot}
	\label{kaimal_volt_probplot}
\end{figure}

\begin{figure}[]
	\centering
	\includegraphics[width=1.00\linewidth]{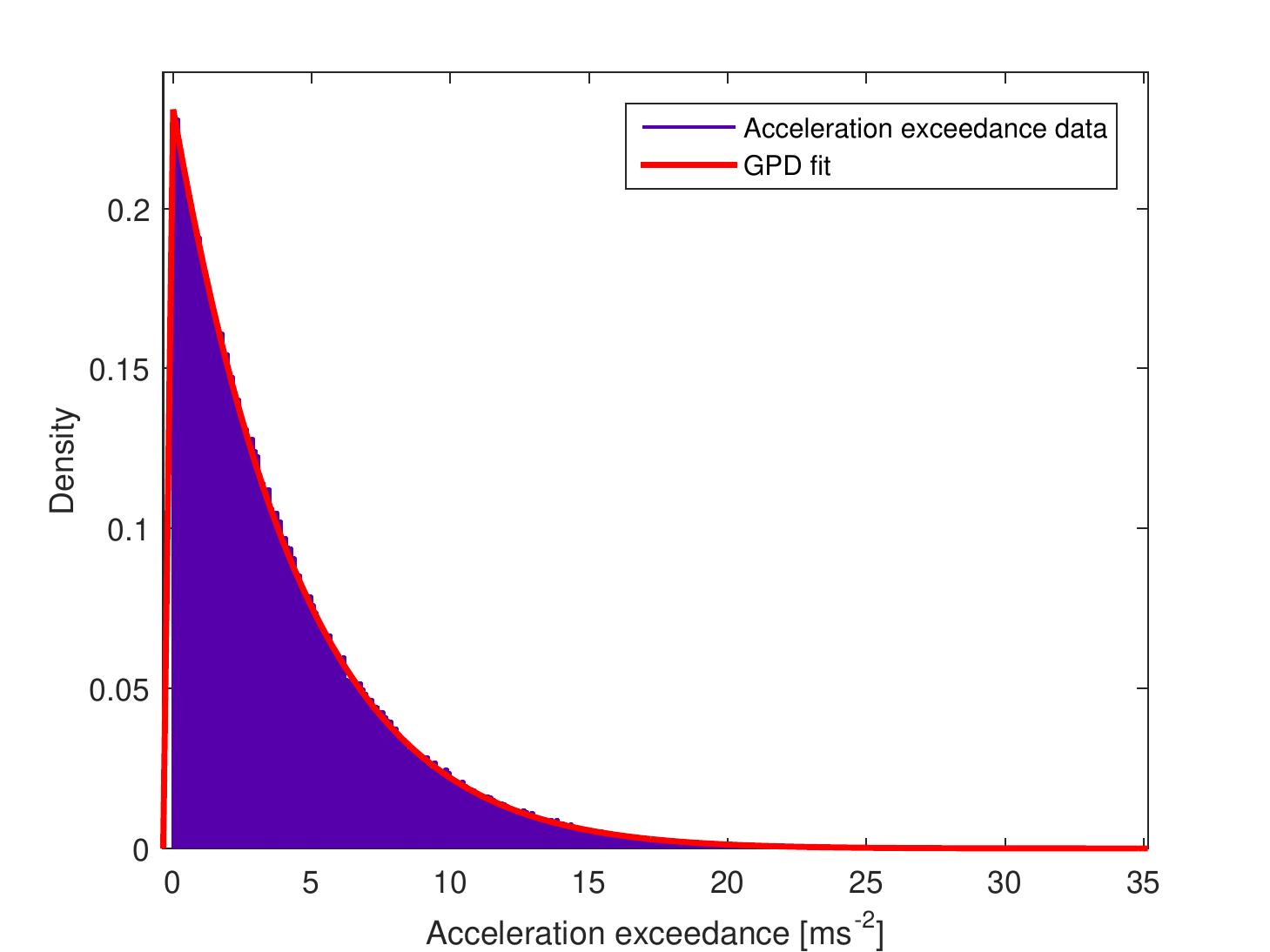}
	\caption{Kaimal acceleration threshold exceedance density plot}
	\label{kaimal_accel_PDF}
\end{figure}

\begin{figure}[]
	\centering
	\includegraphics[width=1.00\linewidth]{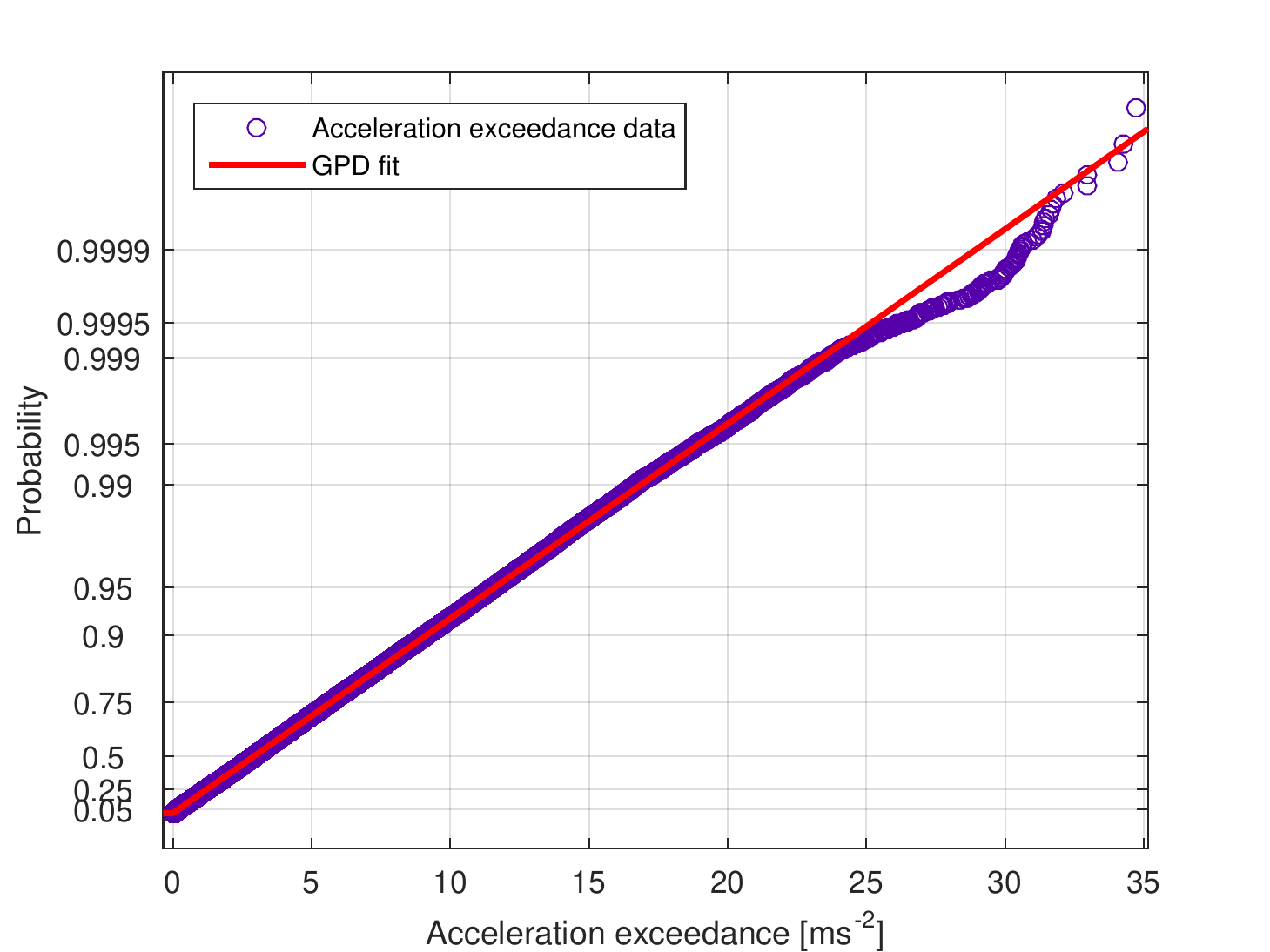}
	\caption{Kaimal acceleration threshold exceedance probability plot}
	\label{kaimal_accel_probplot}
\end{figure}

\begin{figure}[]
	\centering
	\includegraphics[width=1.00\linewidth]{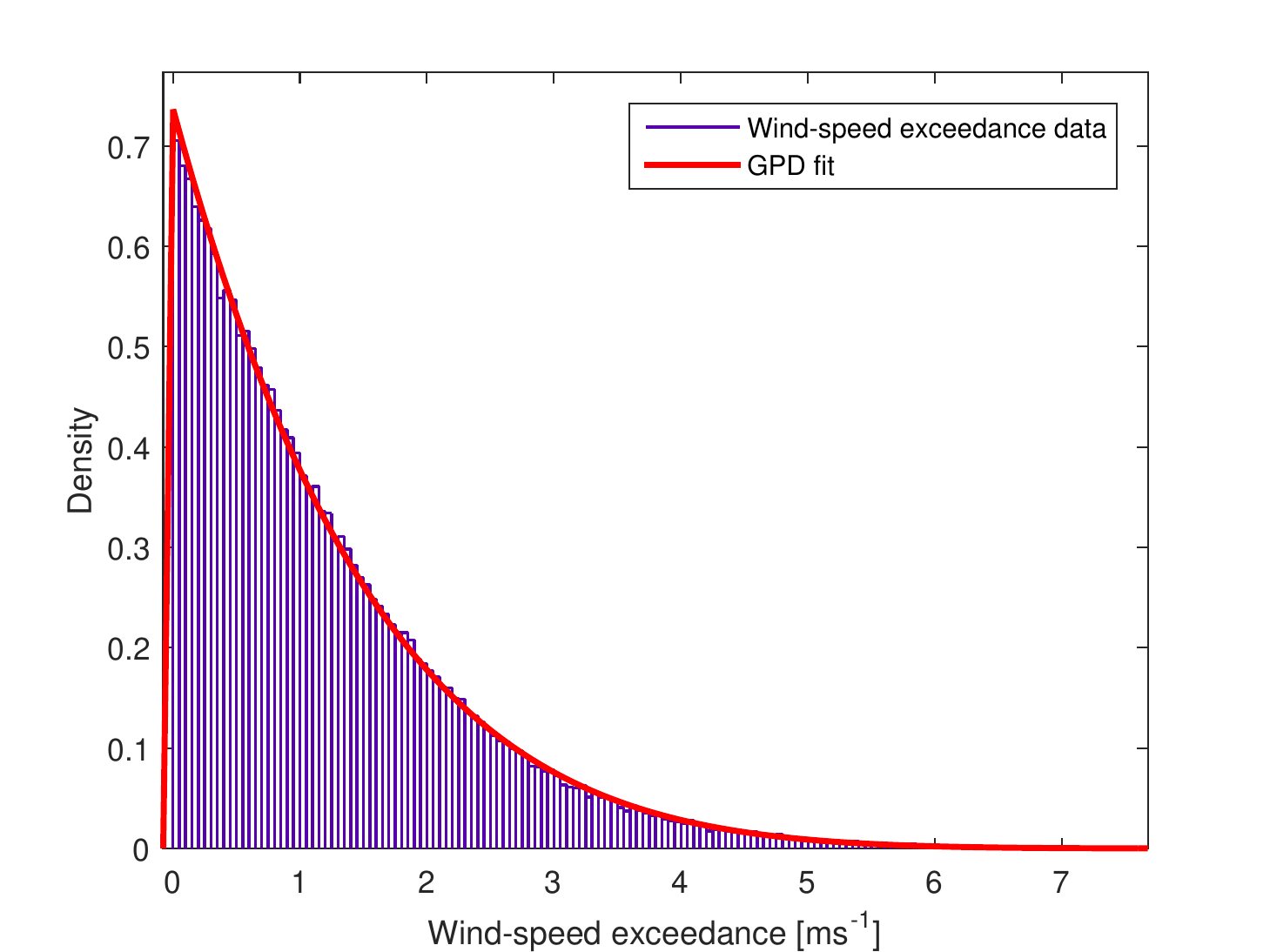}
	\caption{Kaimal wind-speed threshold exceedance density plot}
	\label{kaimal_speed_PDF}
\end{figure}

\begin{figure}[]
	\centering
	\includegraphics[width=1.00\linewidth]{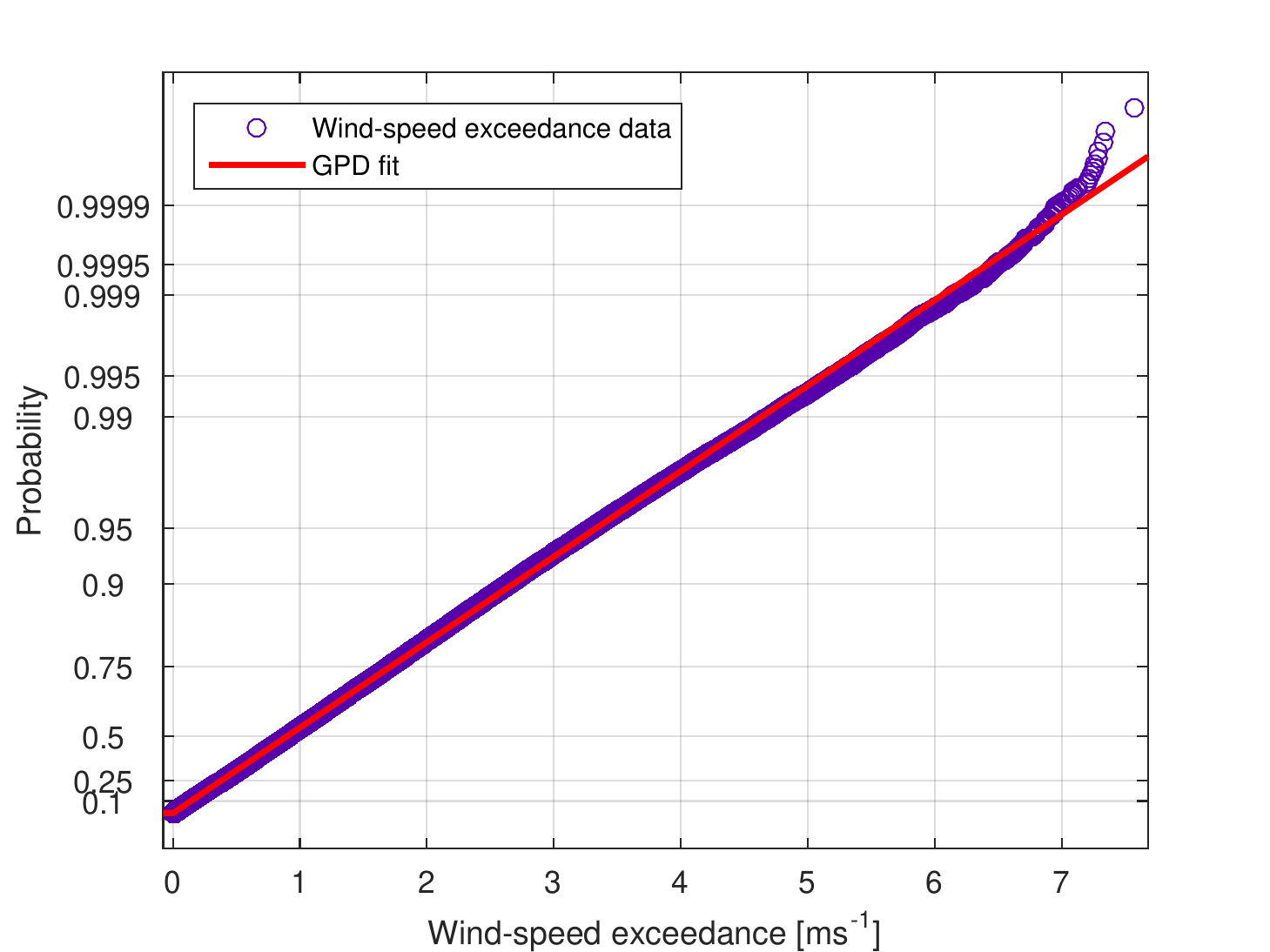}
	\caption{Kaimal wind-speed threshold exceedance probability plot}
	\label{kaimal_speed_probplot}
\end{figure}
\end{subfigures}

\begin{subfigures}
\begin{figure}[]
	\centering
	\includegraphics[width=1.00\linewidth]{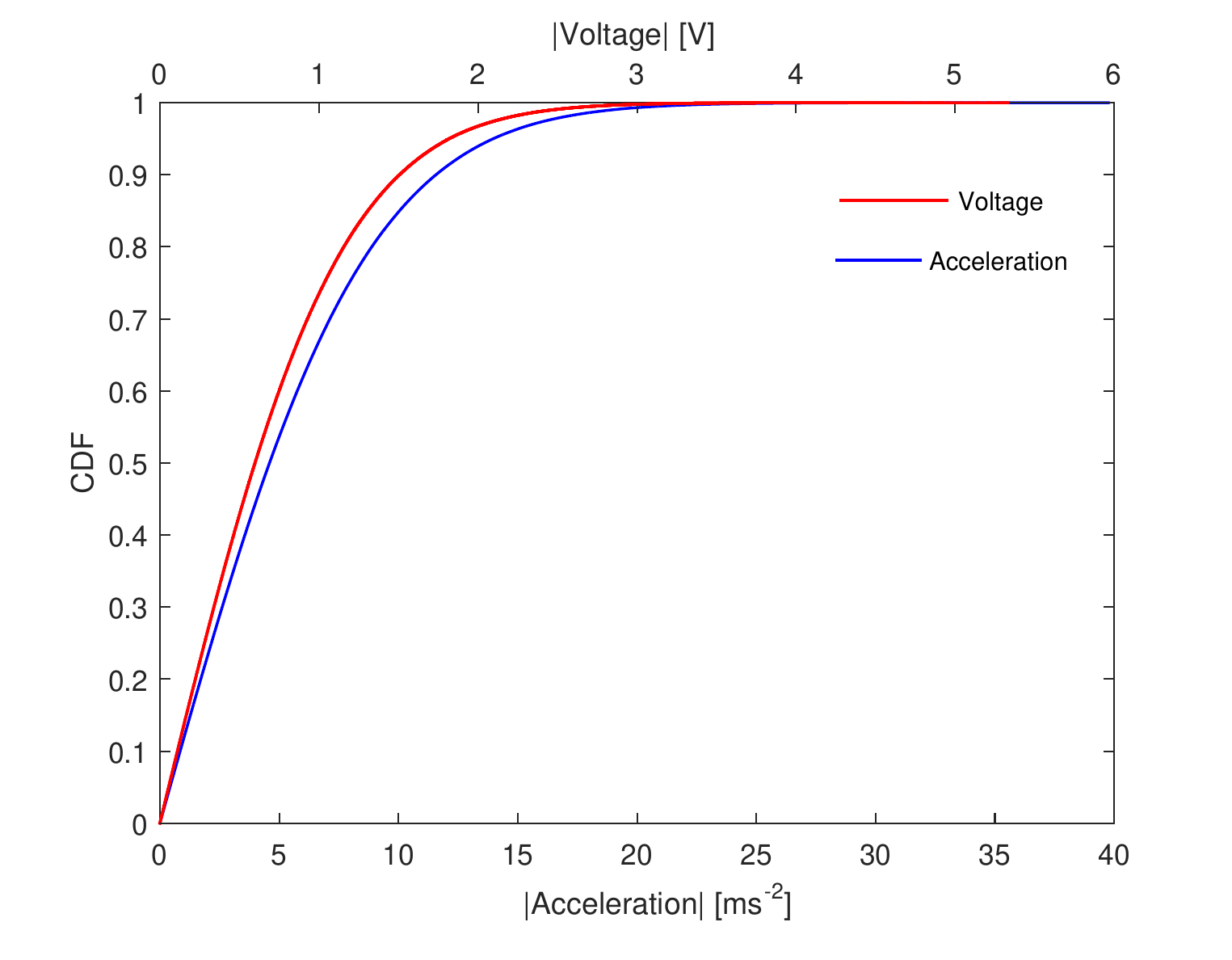}
	\caption{Davenport empirical CDF}
	\label{CDF_combined_davenport}
\end{figure}

\begin{figure}[]
	\centering
	\includegraphics[width=1.00\linewidth]{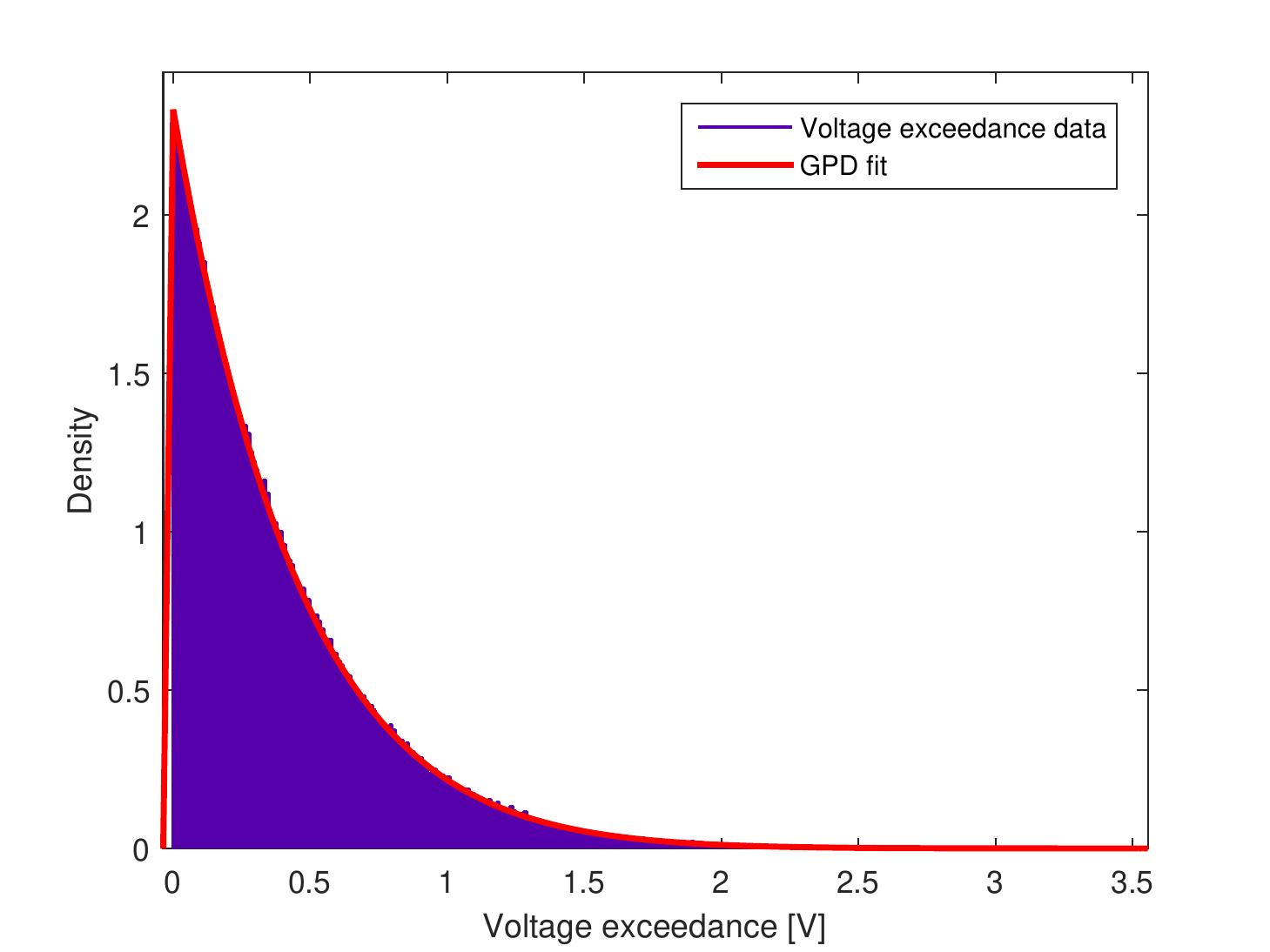}
	\caption{Davenport voltage threshold exceedance density plot}
	\label{davenport_volt_PDF}
\end{figure}
\begin{figure}[]
	\centering
	\includegraphics[width=1.00\linewidth]{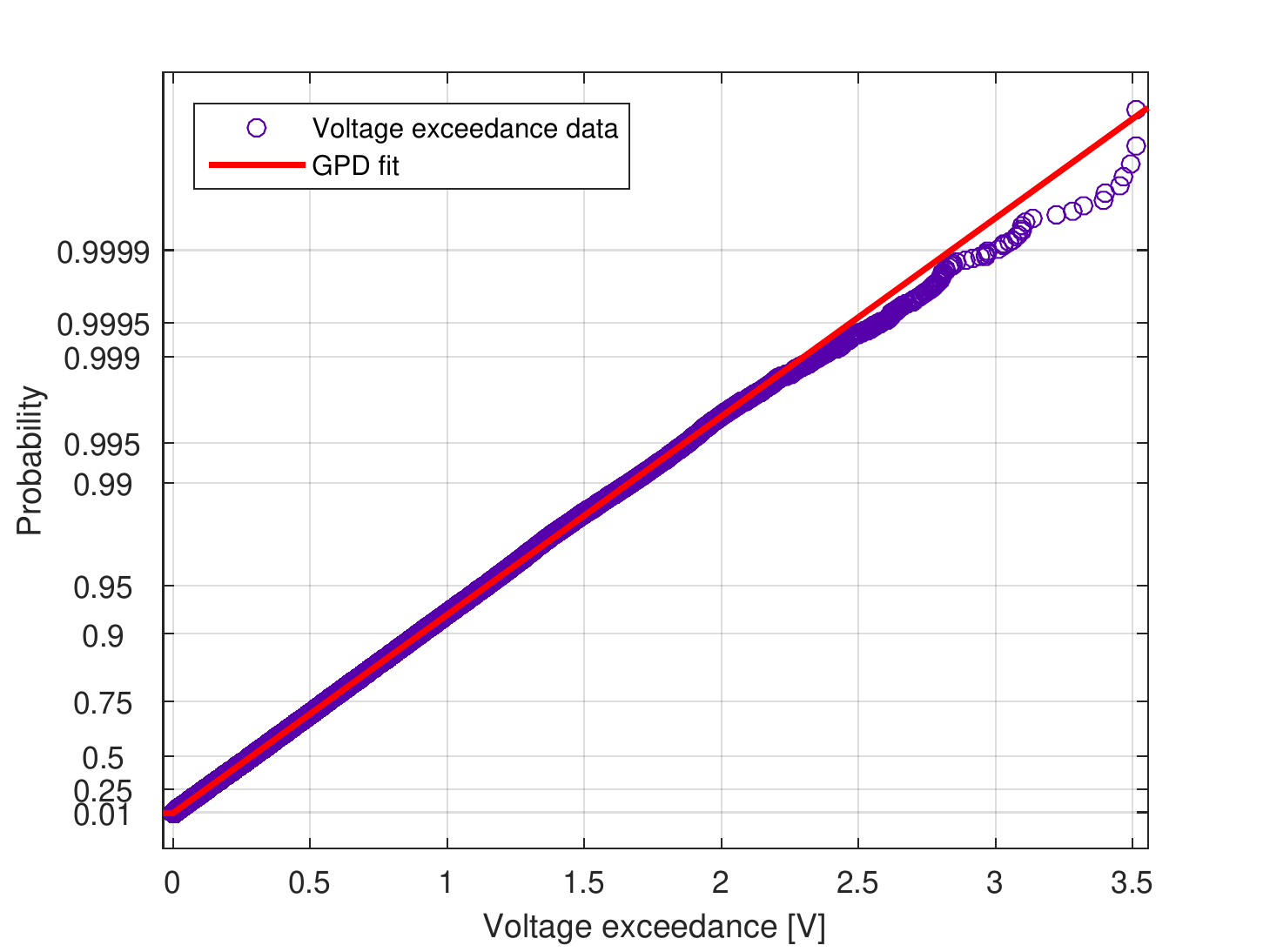}
	\caption{Davenport voltage threshold exceedance probability plot}
	\label{davenport_volt_probplot}
\end{figure}

\begin{figure}[]
	\centering
	\includegraphics[width=1.00\linewidth]{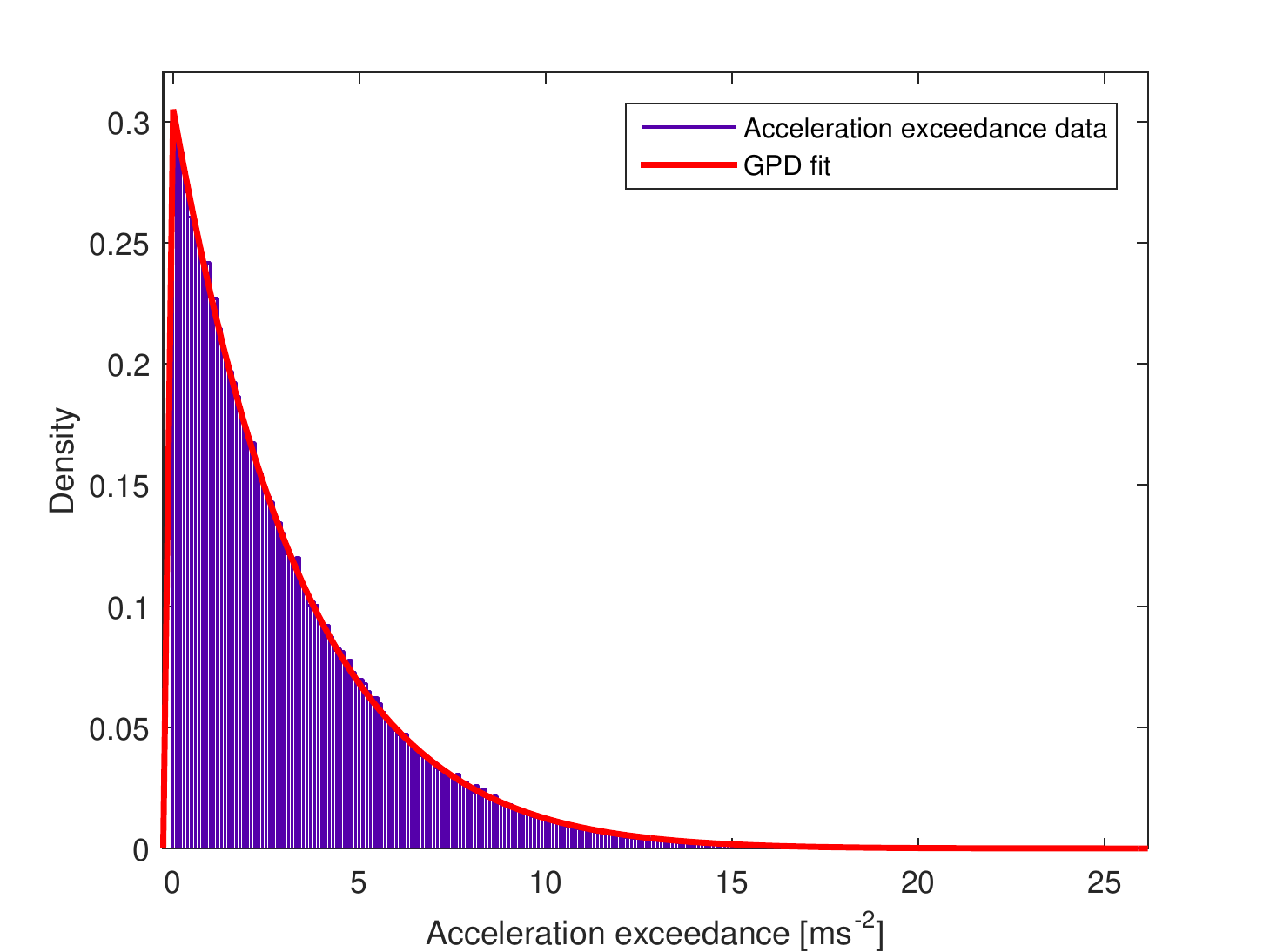}
	\caption{Davenport acceleration threshold exceedance density plot}
	\label{davenport_accel_PDF}
\end{figure}
\begin{figure}[]
	\centering
	\includegraphics[width=1.00\linewidth]{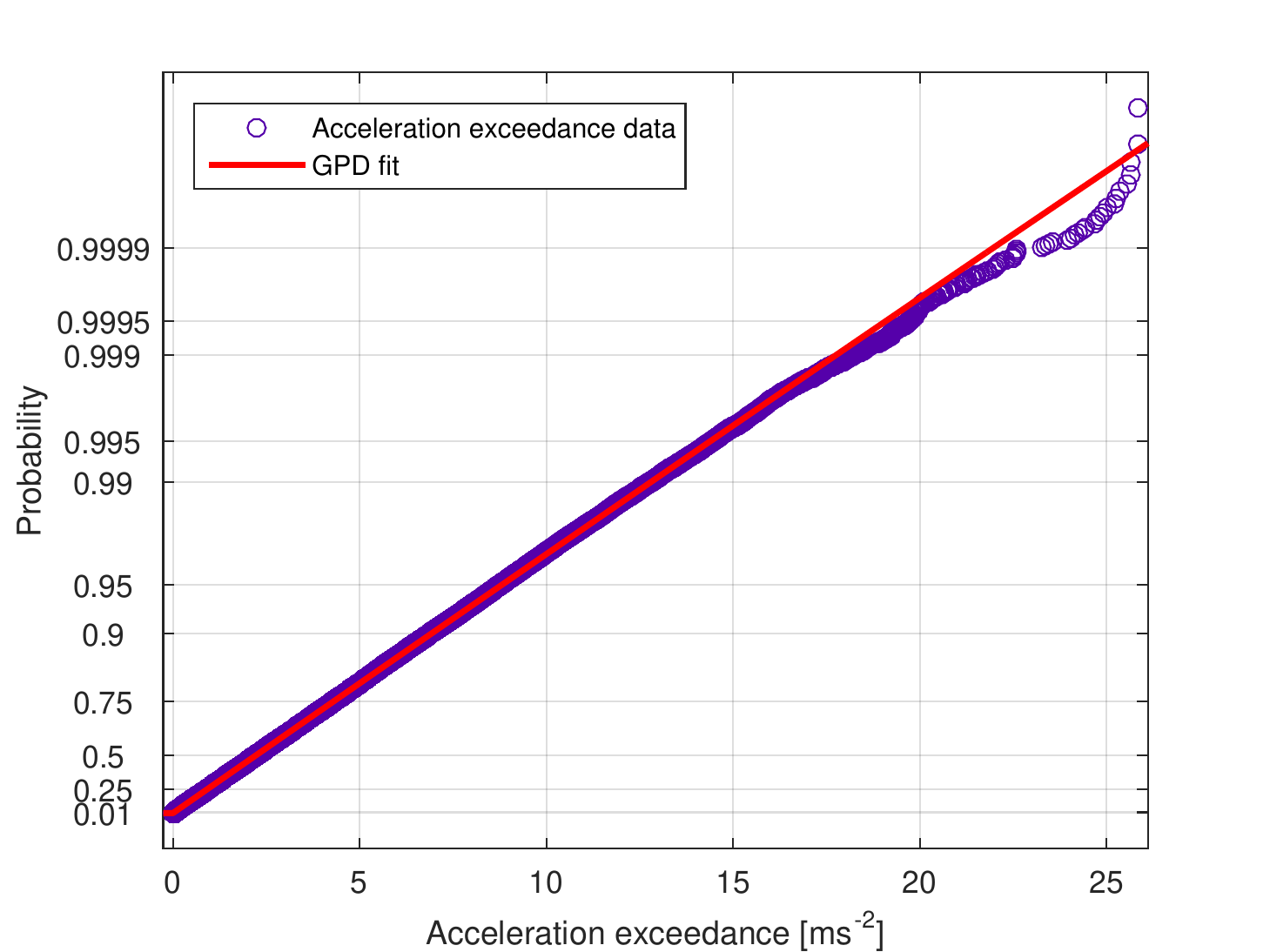}
	\caption{Davenport acceleration thrshold exceedance probability plot}
	\label{davenport_accel_probplot}
\end{figure}
\begin{figure}[]
	\centering
	\includegraphics[width=1.00\linewidth]{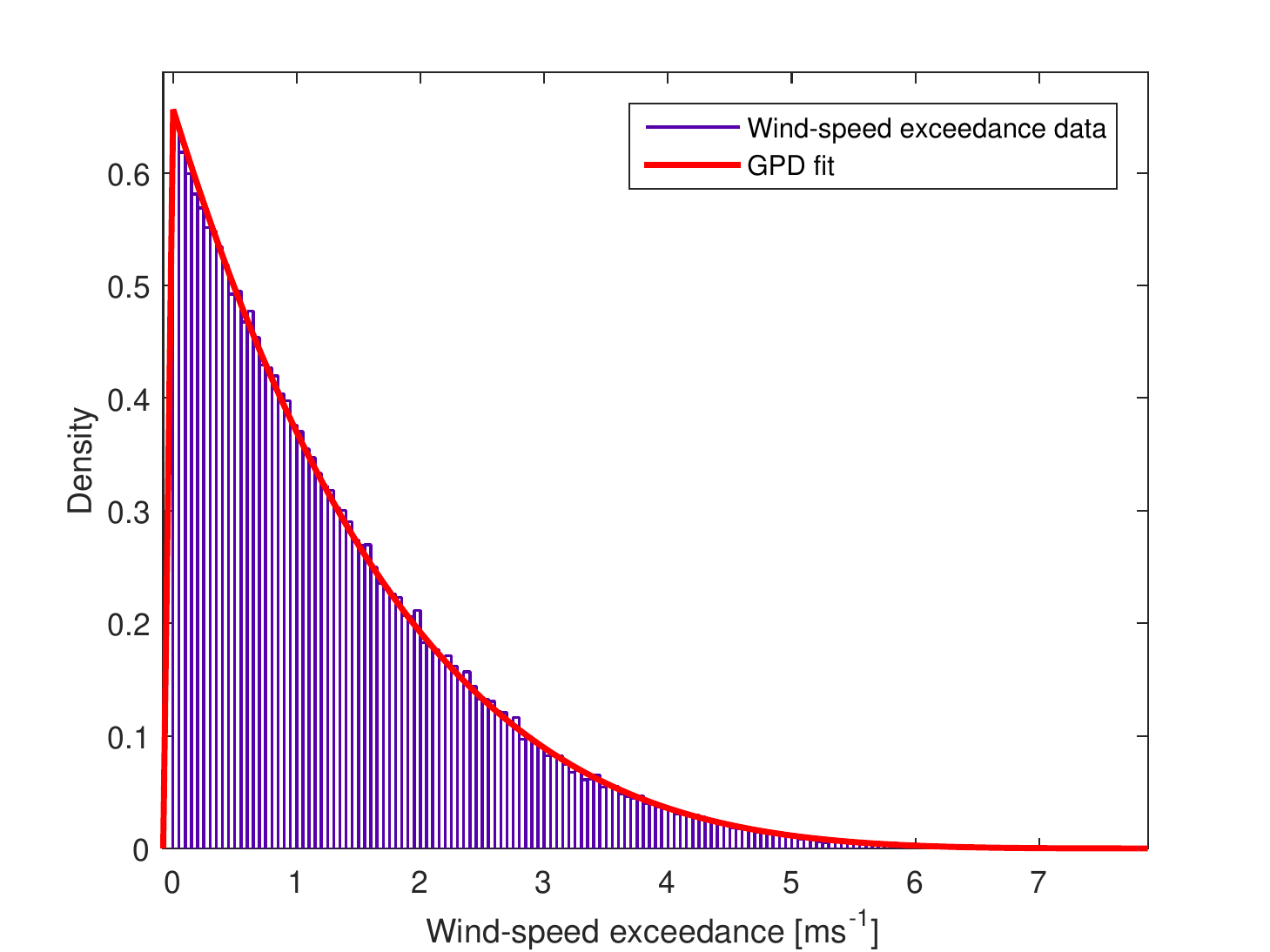}
	\caption{Davenport wind-speed threshold exceedance density plot}
	\label{davenport_speed_PDF}
\end{figure}
\begin{figure}[]
	\centering
	\includegraphics[width=1.00\linewidth]{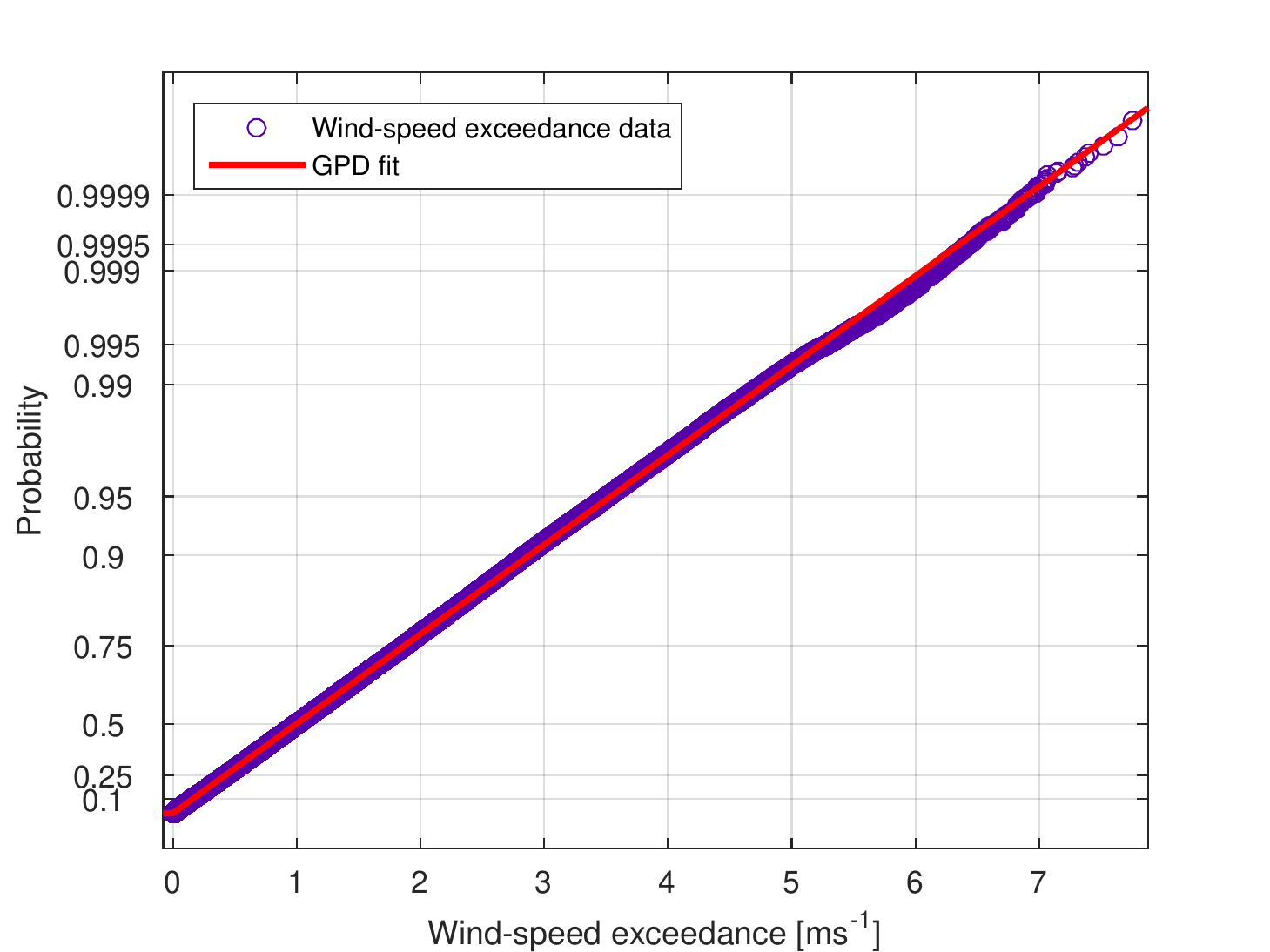}
	\caption{Davenport wind-speed threshold exceedance probability plot}
	\label{davenport_speed_probplot}
\end{figure}
\end{subfigures}

We now proceed to estimate the return levels from our modelled distributions. We are interested in obtaining the threshold exceedance $y$ with a probability of 0.05 or 0.01, or in general $p$. ie.

\begin{linenomath*}
\begin{equation}
Pr(X>u+y\,|\,X>u) = p
\end{equation}
\end{linenomath*}

which can be expressed as $1-$(CDF), ie.

\begin{linenomath*}
\begin{equation}
\label{Prob_CDF}
Pr(X>u+y\,|\,X>u) = \Bigg[1 + \frac{\xi y}{\sigma}\Bigg]^{-1/\xi}
\end{equation}
\end{linenomath*}

Now using Bayes theorem,

\begin{linenomath*}
\begin{equation}
Pr(X>u+y\,|\,X>u) = \frac{Pr(X>u+y)}{Pr(X>u)}
\end{equation}
\end{linenomath*}

giving

\begin{linenomath*}
\begin{equation}
\begin{split} Pr(X>u+y) &= Pr(X>u)\,\Bigg[1 + \frac{\xi y}{\sigma}\Bigg]^{-1/\xi}\\
 &= \lambda\,\Bigg[1 + \frac{\xi y}{\sigma}\Bigg]^{-1/\xi}
\end{split}
\end{equation}
\end{linenomath*}

where $\lambda$ is the empirical threshold exceedance probability. Substituting the raw observation (before calculating exceedance, $y = x-u$) into the equation:

\begin{linenomath*}
\begin{equation}
Pr(X>x) = \lambda\,\Bigg[1 + \frac{\xi (x-u)}{\sigma}\Bigg]^{-1/\xi}
\end{equation}
\end{linenomath*}

Thus, an estimate of the level $z$ that is exceeded on average once every $t$ observations is obtained by solving

\begin{linenomath*}
\begin{equation}
\label{return}
\lambda\,\Bigg[1 + \frac{\xi (z-u)}{\sigma}\Bigg]^{-1/\xi} = \frac{1}{t}
\end{equation}
\end{linenomath*}

giving

\begin{linenomath*}
\begin{equation}
z = u+ \frac{\sigma}{\xi}\Big[\left(t\lambda\right)^\xi-1\Big]
\end{equation}
\end{linenomath*}

$z$ is the $t$-observation return level. If there are n observations per year, then the $r$-year return level can be expressed as

\begin{linenomath*}
\begin{equation}
z = u+ \frac{\sigma}{\xi}\Big[\left(rn\lambda\right)^\xi-1\Big]
\end{equation}
\end{linenomath*}

Now, for the same number of observations, equation \ref{return} should be the same for wind-speed, acceleration and voltage. ie., 

\begin{linenomath*}
\begin{equation}
\begin{split}
\lambda_w\,\Bigg[1 + \frac{\xi_w (z_w-u_w)}{\sigma_w}\Bigg]^{-1/\xi_w} =\lambda_a\,\Bigg[1 + \frac{\xi_a (z_a-u_a)}{\sigma_a}\Bigg]^{-1/\xi_a}\\ =\lambda_v\,\Bigg[1 + \frac{\xi_v (z_v-u_v)}{\sigma_v}\Bigg]^{-1/\xi_v}
\end{split}
\end{equation}
\end{linenomath*}

Using the same empirical threshold exceedance probability, $\lambda$, for all three data-sets gives us the mapping between the return levels as:

\begin{linenomath*}
\begin{equation}
\begin{split}
\Bigg[1 + \frac{\xi_w (z_w-u_w)}{\sigma_w}\Bigg]^{-1/\xi_w} =\Bigg[1 + \frac{\xi_a (z_a-u_a)}{\sigma_a}\Bigg]^{-1/\xi_a}\\ =\Bigg[1 + \frac{\xi_v (z_v-u_v)}{\sigma_v}\Bigg]^{-1/\xi_v}
\end{split}
\end{equation}
\end{linenomath*}

\begin{figure}[!htb]
	\centering
	\includegraphics[width=1.00\linewidth]{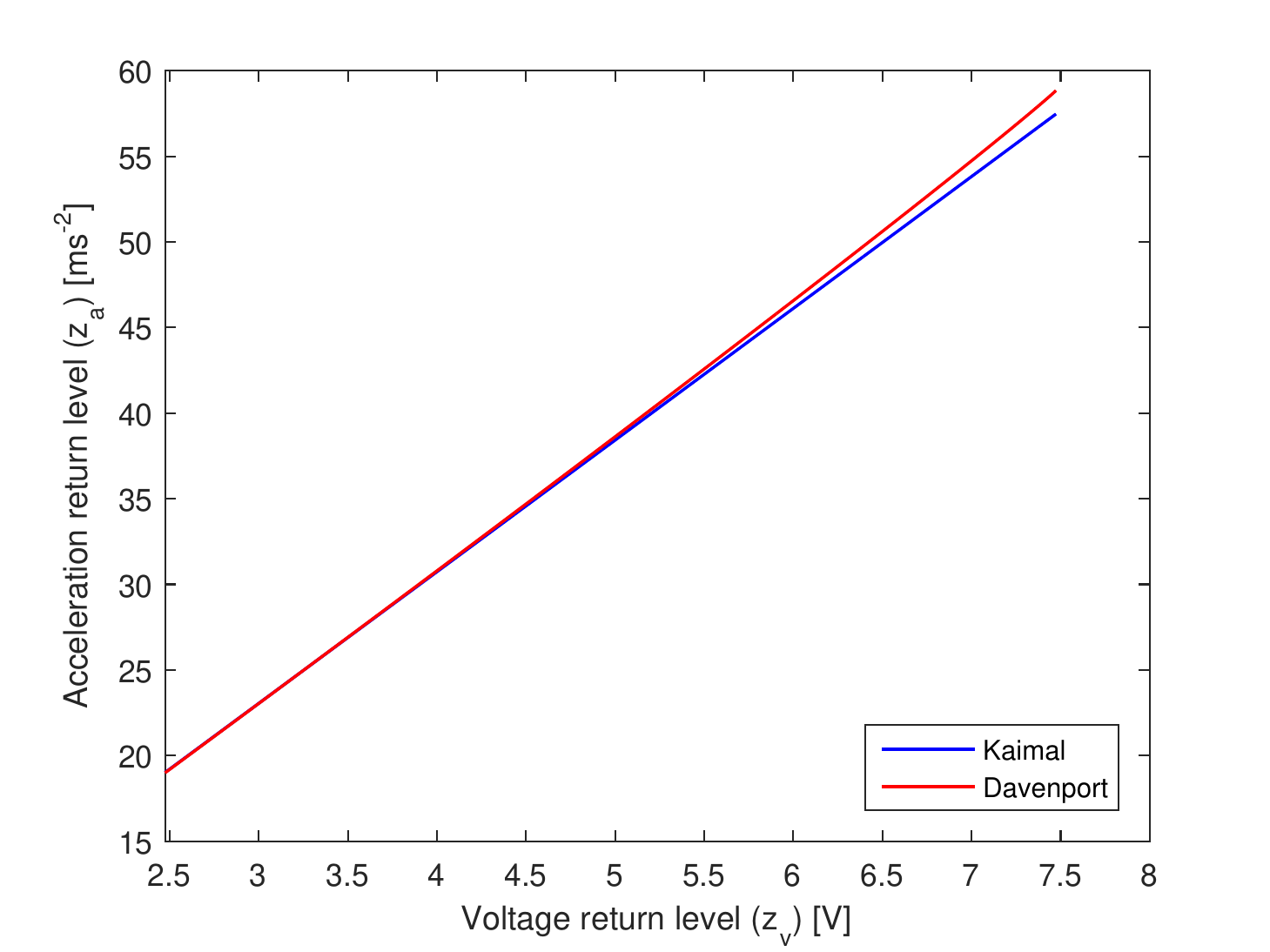}
	\caption{Acceleration return level as a function of voltage return level}
	\label{returnfigaccel}
\end{figure}

\begin{figure}[!htb]
	\centering
	\includegraphics[width=1.00\linewidth]{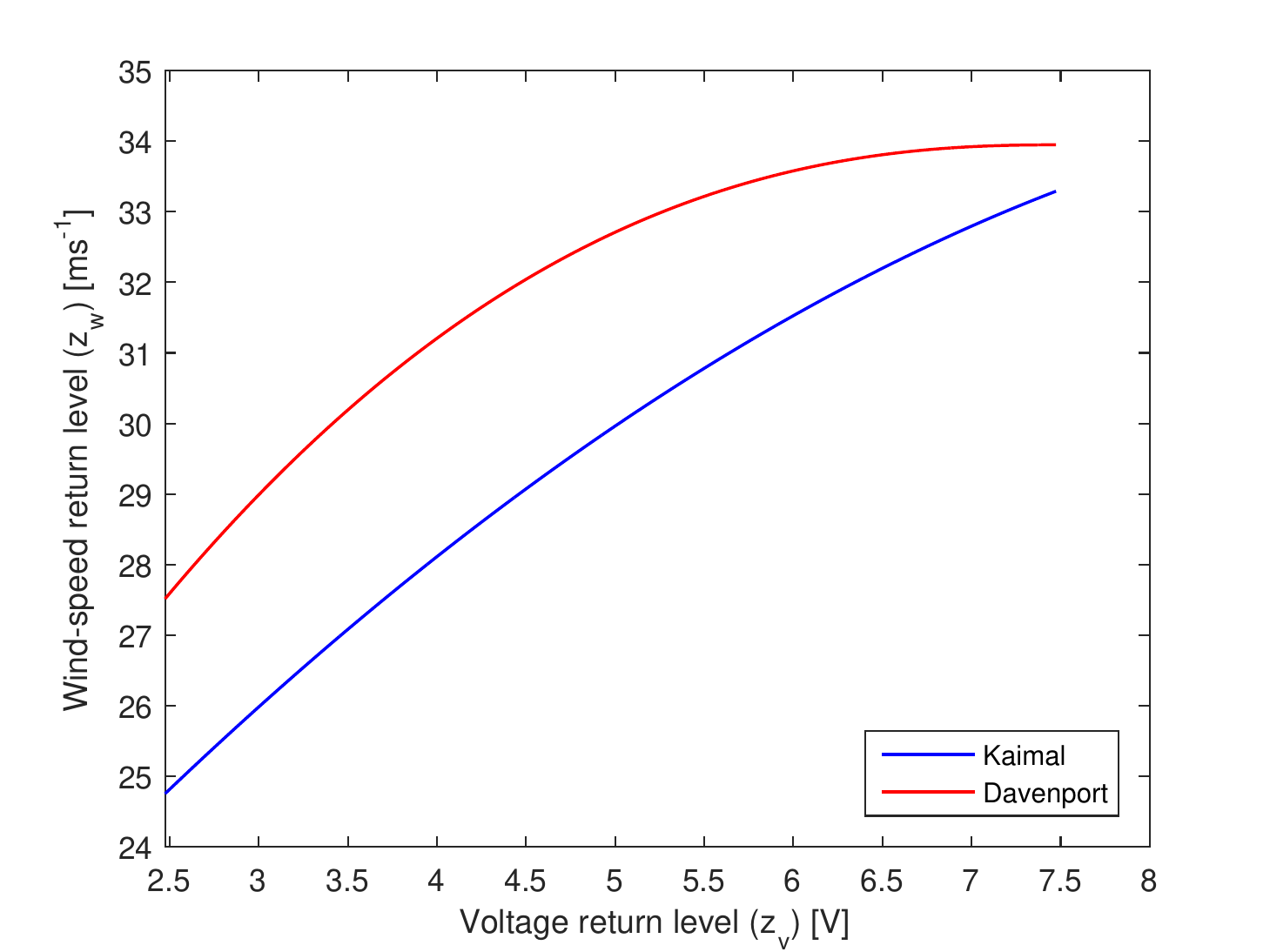}
	\caption{Wind-speed return level as a function of voltage return level}
	\label{returnfigwind}
\end{figure}

Once calibrated using an appropriate GPD model, the return levels of the acceleration($z_a$; Eq.\ref{returnacceleq}, Fig.\ref{returnfigaccel}) of the host structure, and wind-speed($z_w$; Eq.\ref{returnwindeq}, Fig.\ref{returnfigwind}) can be expressed as functions of the return level of voltage($z_v$). 

\begin{linenomath*}
\begin{equation}
\label{returnacceleq}
z_a = u_a + \frac{\sigma_a}{\xi_a}\,\Bigg[\Bigg(1 + \frac{\xi_v(z_v-u_v)}{\xi_v} \Bigg)^{\frac{\xi_a}{\xi_v}} - 1 \Bigg]
\end{equation}
\end{linenomath*}

\begin{linenomath*}
\begin{equation}
\label{returnwindeq}
z_w = u_w + \frac{\sigma_w}{\xi_w}\,\Bigg[\Bigg(1 + \frac{\xi_v(z_v-u_v)}{\xi_v} \Bigg)^{\frac{\xi_w}{\xi_v}} - 1 \Bigg]
\end{equation}
\end{linenomath*}

The number of iterations employed and thereby the size of the data-set is relatively small compared to raw-data sets sometimes used for extreme-value analysis \cite{naess1998estimation,caires2005100}. An increase in the number of iterations in this study would only benefit towards improved proximity to the convergent asymptotic values. The work is thus carried out within the bounds of situations where Eq.\ref{Prob_CDF} is true, which is valid for a wide range of cases. The method of calibration is of primary interest over the values estimated.

The benchmark study using Gaussian white noise presents a typical excitation of SDOF oscillator and harvester. The voltage and acceleration is expected to be Gaussian as the cascaded systems are linear, and it is indeed. The forcing function due to wind would be close to a chi-squared distribution as the wind-speed time-series is generated here using a Gaussian distribution. The Gaussian-ness of the resultant acceleration and voltage may not be obvious in this case. It is due to the bandwidth of the transfer function of the SDOF oscillator and harvester. For a non-Gaussian forcing function that has a wide bandwidth compared with the bandwidth of the system transfer function, for most cases, the output response will be Gaussian\cite{mazelsky2012extension}.


\section{Conclusion}
This paper has established how analysis of harvested energy from vibrations of structures can be used to estimate extreme value responses and related return periods. A cantilever-style piezoelectric harvester connected to an SDOF system was considered in this regard, excited by wind loading derived from Kaimal and Davenport spectra respectively, in addition to Gaussian white noise excitation. The extremes were fitted using an asymptotic Generalized Pareto Distribution. The mapping was calibrated for wind-loading indicating return levels of wind-speed as a function of return levels of harvested voltage. The findings extend the capabilities of energy harvesters as monitors of built infrastructure and can be useful for both researchers and professionals.
\section*{Acknowledgements}
The authors would like to acknowledge J.N. Tata Endowment and Marine and Renewable Energy Ireland (MaREI), grant no. 12/RC/2302, a Science Foundation Ireland (SFI) supported project.








\end{document}